# Optimisation de la QoS dans un réseau de radio cognitive en utilisant la métaheuristique SFLA (Shuffled Frog Leaping Algorithm)

Badr Benmammar

badr.benmammar@gmail.com
Laboratoire de Télécommunications Tlemcen
Université de Tlemcen, Algérie

**Résumé :**

Ce rapport propose une étude de la qualité de service (QoS) dans le contexte de la radio cognitive. Cette étude est basée sur une méthode d’optimisation stochastique appelée algorithme de saut de grenouille (shuffled frog leaping algorithm). L’intérêt de l’algorithme SFLA est de garantir une meilleure solution dans un contexte multi porteuses afin de satisfaire les exigences de l’utilisateur secondaire (SU).

**Abstract :**

This paper proposes a study of quality of service (QoS) in cognitive radio networks. This study is based on a stochastic optimization method called shuffled frog leaping algorithm (SFLA). The interest of the SFLA algorithm is to guarantee a better solution in a multi-carrier context in order to satisfy the requirements of the secondary user (SU).

# Table des matières

# Introduction générale

De plus en plus, les besoins de communication ne cessent d'augmenter, que ce soit pour des besoins de contrôle, de vidéo surveillance ou de nouveaux services offerts à la clientèle. Ce qui a conduit à des problèmes de cohabitation et d'interopérabilité entre systèmes et de disponibilité des fréquences. d'autre part, ces systèmes doivent garantir des exigences de disponibilité, de continuité de service, d'hétérogénéité des trafics, de robustesse et de qualité de service des applications visées parfois dans un contexte de forte mobilité et quels que soient les environnements traversés (rural, tranchées, tunnels…). Certaines normes émergentes (IEEE.802.16m, IEEE.802.20, A-LTE) [1], basées sur la technologie MIMO et celle de multi-porteuses OFDM [2], pourraient répondre aux besoins des ITS (Intelligent Transport Systems) en terme de débit [3]. Alors que les travaux développés actuellement sont généralement basés sur l'emploi d'antennes simples en émission/réception, de nouvelles approches doivent être envisagées afin d'intégrer un schéma à antennes multiples MIMO-OFDM. En effet, la gestion statique du spectre de fréquence radio qui caractérise actuellement les réseaux de radiocommunication traditionnels conduit à une pénurie de bandes radio disponibles (i.e. une large proportion du spectre alloué n'est utilisée que de manière sporadique).

La solution pour l'amélioration de la gestion et de l'utilisation du spectre radio est d'appliquer la technologie de la radio cognitive (RC). Les principales techniques qui permettent de tirer le maximum de bénéfice de cette technologie sont l'allocation efficace de ressources radio et l'analyse de l'environnement radio [4].

Dans le contexte de ce travail, nous nous intéressons aux réseaux de radio cognitive et à l'accès dynamique au spectre. En particulier nous cherchons des solutions pour une meilleure utilisation du spectre et une amélioration de la qualité de service fournie à l'utilisateur.

L'objectif de notre travail est de résoudre le problème de l'accès dynamique au spectre et d'optimiser la qualité de service (QoS) en utilisant la méta-heuristique Suffled Frog Leaping Algorithme.

Ce rapport est structuré comme suit :

- Chapitre I : Réseaux de radio cognitive

  Ce chapitre présente un ensemble de notions importantes concernant la radio cognitive, ainsi que ses principes, ses objectifs et ses domaines d'application.

- Chapitre II : Les métaheuristiques

  Cette partie aborde l'heuristique et les métaheuristiques en citant leur notions et classification qui se compose de deux grandes classes : Métaheuristique à solution unique et Métaheuristique à solution multiple. Chaque classe comprend un ensemble de méthodes. Par suite, l'algorithme SFLA (Shuffled Frog Leaping) va être présenté avec ses étapes et son organigramme.

- Chapitre III : Evaluation et test

  Ce dernier chapitre est consacré à l'évaluation de notre proposition en indiquant la fonction objectif, les modes de transmission utilisés, les expérimentations et les résultats obtenus. Enfin, notre algorithme SFLA va être comparé avec les algorithmes génétiques en termes de fitness.

# Chapitre I : Réseau de radio cognitive

## I.1 Introduction

La rapidité d'évolution des technologies sans fil entraine une forte demande en termes de ressources spectrales. Pour résoudre ce problème il faut une bonne gestion du spectre et donc une utilisation plus efficace et une exploitation opportuniste de celui-ci. En effet, il existe des bandes inutilisables desquelles il est possible de profiter pour augmenter le nombre d'utilisateurs et pour mieux répartir les ressources disponibles [5].

Une telle aptitude relève du concept de radio cognitive (RC) introduit par Joseph Mitola en 2000 [6]. Un terminal RC peut donc interagir avec son environnement radio afin de s'y adapter, d'y détecter les fréquences libres et de les exploiter. Le terminal aura suffisamment les capacités lui permettant de gérer efficacement l'ensemble des ressources radio. Les recherches actuelles sur la radio cognitive s'intéressent principalement à la détection des ressources libres et à la répartition dynamique des fréquences entre les terminaux RC [7].

Nous allons présenter dans ce chapitre la radio cognitive dans ses différents aspects : la gestion spectrale, principes, architecture et application passant par les différentes radios.

## I.2 Gestion spectrale

Les bandes spectrales inutilisables ont des caractéristiques différentes les unes des autres. Ces caractéristiques sont la fréquence d'opération de la bande spectrale, le débit et le temps. Toutes ces informations changent au cours du temps vu la nature dynamique de l'environnement radio. Les nouvelles fonctions qui sont présentées par (Akyildiz et al) [8] sont requises pour gérer les ressources spectrales dans les RRC (Réseaux de Radio Cognitive). Ces fonctions sont la détection spectrale, l'analyse spectrale et la décision spectrale.

### I.2.1 Détection spectrale

Détecter le spectre non utilisé et le partager sans interférence avec d'autres utilisateurs. La détection des utilisateurs primaires est la façon la plus efficace pour détecter les espaces blancs du spectre. L'un des objectifs de la détection du spectre, en particulier pour la détection des interférences, est d'obtenir le statut du spectre (libre /occupé), de sorte que le spectre peut être consulté par un utilisateur secondaire en vertu de la contrainte d'interférence. Le défi réside dans le fait de mesurer l'interférence au niveau du récepteur primaire causé par les transmissions d'utilisateurs secondaires [9].

### I.2.2 Analyse spectrale

Elle permet de caractériser les différentes bandes spectrales en termes de fréquence d'opération, de débit, de temps et de l'activité du PU (Utilisateur Primaire). Cette caractérisation sert à répondre aux exigences de l'URC (Utilisateur à Radio Cognitive). Des paramètres supplémentaires viennent compléter cette caractérisation, à savoir, le niveau d'interférence, le taux d'erreur du canal, les évanouissements, le délai et le temps d'occupation de la bande spectrale par un URC.

### I.2.3 Décision spectrale

Après que toutes les bandes spectrales aient été catégorisées et classifiées, on applique un ensemble de règles décisionnelles pour obtenir la ou les bandes spectrales les plus appropriées à la transmission en cours, en tenant compte des exigences de l'URC.

## I.3 Accès dynamique au spectre

Le principe de l'accès dynamique au spectre est d'arriver à une utilisation plus efficace des fréquences. L'existence de fréquences non utilisées dans quelques bandes - des "white spaces" ou "spectrum holes" a justement été signalée à plusieurs reprises. En effet, l'utilisateur primaire n'utilise pas toujours et partout les fréquences qui lui ont été assignées. Le DSA (Dynamic Spectrum Access) permet ici l'implication d'un utilisateur secondaire en mesurant l'occupation du spectre (Spectrum sensing) ou en interrogeant une base de données sur internet, celui-ci peut déterminer si des fréquences sont disponibles à un endroit donné pour une utilisation secondaire. [10]

Donc, deux principaux types d'acteurs dans un réseau à radio cognitive sont distingués : des utilisateurs secondaires et des utilisateurs primaires.

- Utilisateur Secondaire (Secondary User ou SU) : c'est l'utilisateur à radio cognitive qui ne possède pas de licence et souhaite utiliser les parties libres du spectre.
- Utilisateur Primaire (Primary User ou PU) : c'est l'utilisateur qui possède une licence pour opérer dans une bande de fréquence bien spécifique. Il est donc prioritaire pour l'utilisation de cette bande de spectre.

Les SU peuvent accéder aux bandes de spectre inoccupées durant l'absence des PU et libèrent le canal dès qu'un PU veut y accéder mais peuvent également continuer à utiliser ces bandes sous réserve d'obtention de l'accord du PU.

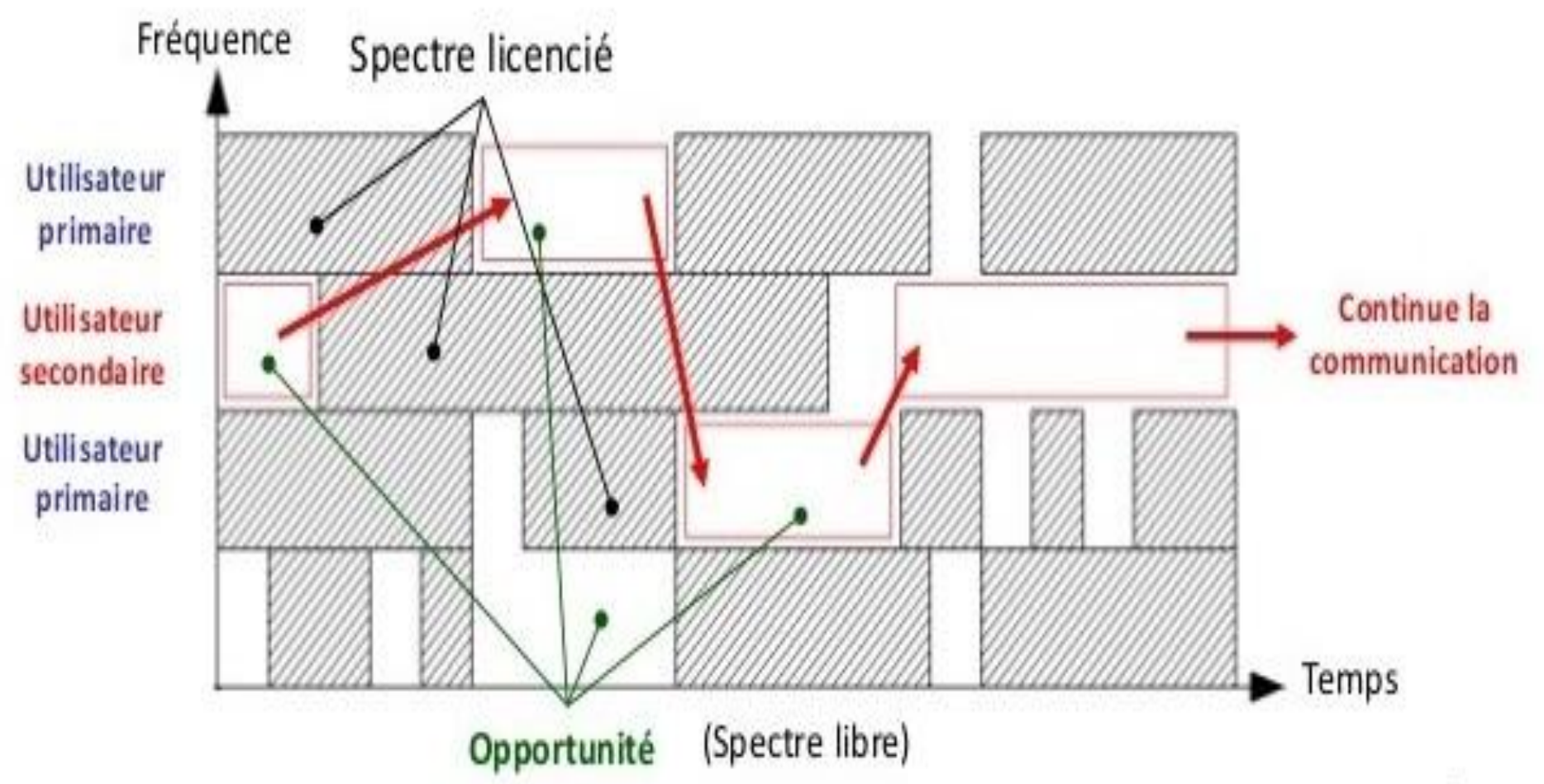

**Figure I. 1:** Exemple d'accès dynamique au spectre [11]

Pour qu'une radio cognitive (utilisateur secondaire) établie une communication sur un réseau licencié (primaire) il observe l'activité de transmission des utilisateurs primaires pour prendre une décision quel spectre choisir en se basant sur certaines connaissances (disponibilité, coopération de l'utilisateur primaire, taux d'erreur, etc.).

## I.4 Radio logicielle

La radio logicielle permet idéalement, à des équipements de communiquer avec n'importe quel standard de radiocommunications par la seule modification du logiciel embarqué, et donc sans modification d'un quelconque élément matériel. Cependant, le caractère statique des protocoles actuels de communication pose les questions de l'optimisation de l'efficacité spectrale et de la flexibilité du domaine radio. De cette réflexion, concernant directement la pérennité des télécommunications modernes, est née le domaine de la radio intelligente ou radio cognitive. Cette évolution, aujourd'hui incontournable dans le monde des radiocommunications, donne la possibilité aux appareils de communication, devenus plus autonomes, de choisir les meilleures conditions de communication.

## I.5 Différence entre radio adaptative, cognitive et intelligente

### I.5.1 Radio adaptive

Une radio adaptative peut contrôler sa propre performance ainsi que modifier les paramètres associés à la communication afin de s'adapter en permanence et de choisir le lien le plus efficace de communication en termes de QoS (Quality of Service).

### I.5.2 Radio cognitive

Les radios cognitives fournissent une étape supplémentaire dans la complexité par rapport aux radios adaptatives. Une radio cognitive est consciente de son environnement et de l'état de fonctionnement (par exemple la localisation, l'utilisation du spectre RF et les réglementations locales).

### I.5.3 Radio intelligente

Les radios intelligentes sont des radios cognitives qui font de l'auto-apprentissage (souvent liée à la notion de "machine learning"). L'utilisateur de ce type de radio a la possibilité de prendre des décisions fondées sur l'expérience.

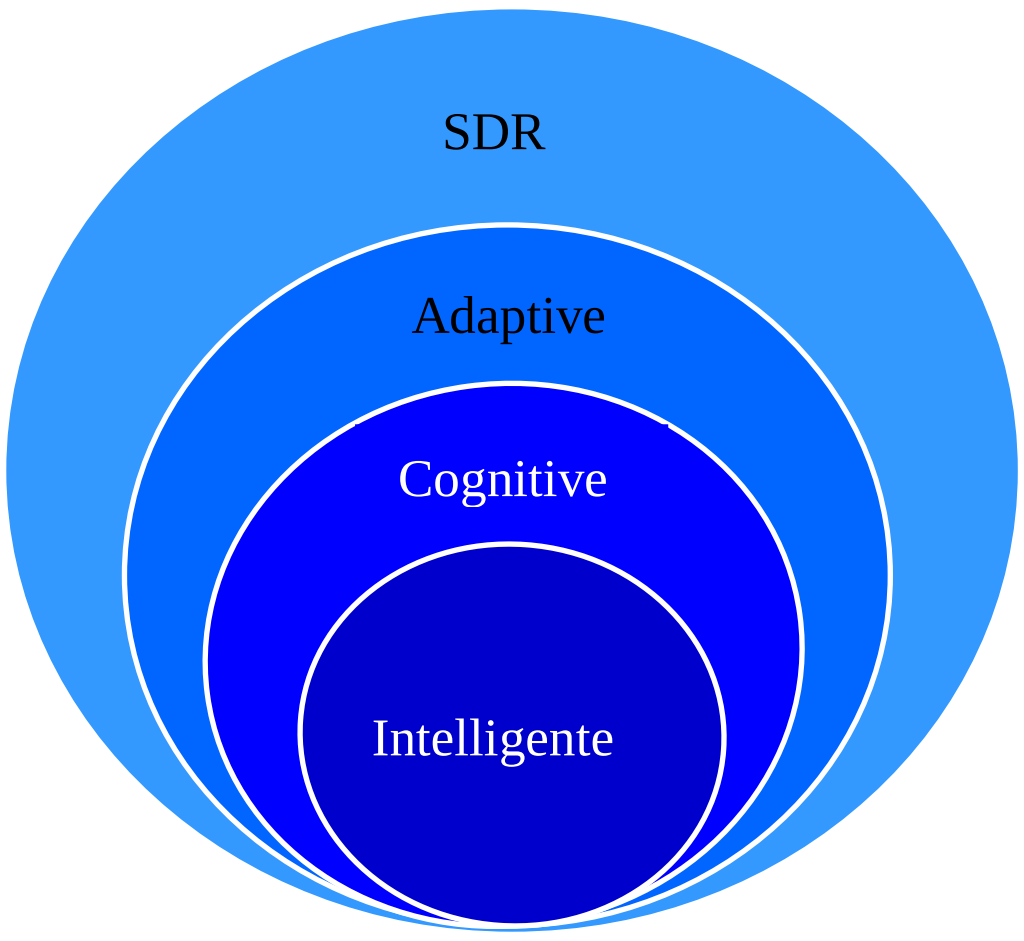

**Figure I. 1:** La relation entre les radios [12]

## I.6 Radio cognitive

### I.6.1 Définition

La Radio cognitive est une forme de communication sans fil dans laquelle un émetteur/récepteur est capable de détecter intelligemment les canaux de communication qui sont en cours d'utilisation et ceux qui ne le sont pas, et peut se déplacer vers les canaux inutilisables. Ceci permet d'optimiser l'utilisation des fréquences radio disponibles du spectre tout en minimisant les interférences avec d'autres utilisateurs [13].

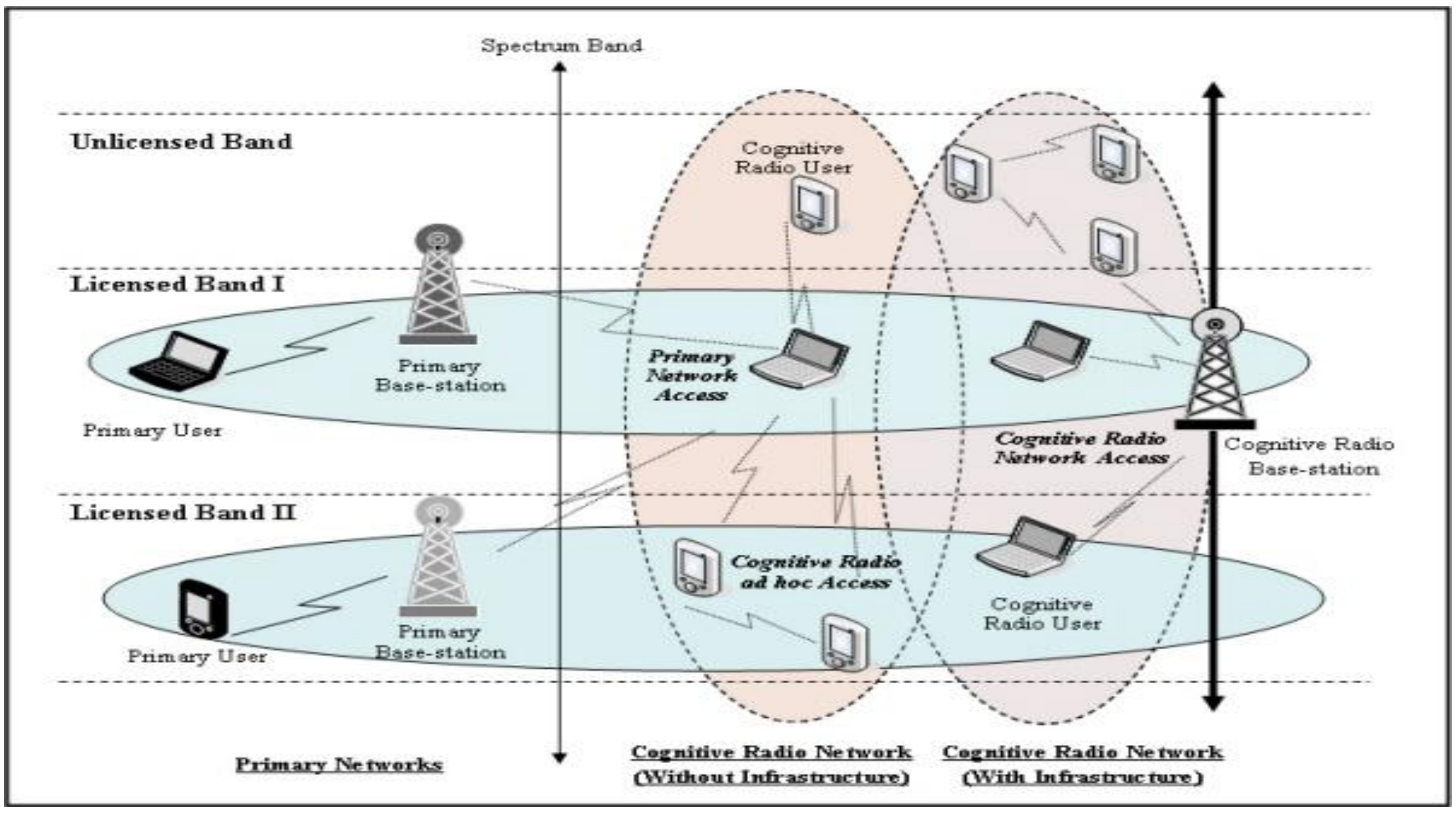

**Figure I.1:** Exemple d'un réseau de radio cognitive [14]

## I.6.2 Radio cognitive (RC) et radio logicielle (SDR)

Dans la radio logicielle SDR (Software Defined Radio) certaines fonctions (filtrage, modulation, etc) sont essentiellement effectuées par la programmation DSP (Digital signal Processor), FPGA (Fields Programmable Gate Array ) , etc. [15].

Le but recherché est d'offrir une alternative plus flexible aux systèmes purement matériels afin d'adapter les transmissions. En conséquence, certaines études relèvent les opportunités qu'il y a à associer la SDR au concept plus général de la RC.

La relation entre la RC et la SDR est définie par un modèle simple qui est représenté dans la figure I.4. Dans ce modèle, les éléments de la RC entourent le support SDR et le cognitive engine qui est nécessaire pour la prise de décision et l'apprentissage de l'environnement radio dans un nuage qui exploite efficacement les ressources disponibles.

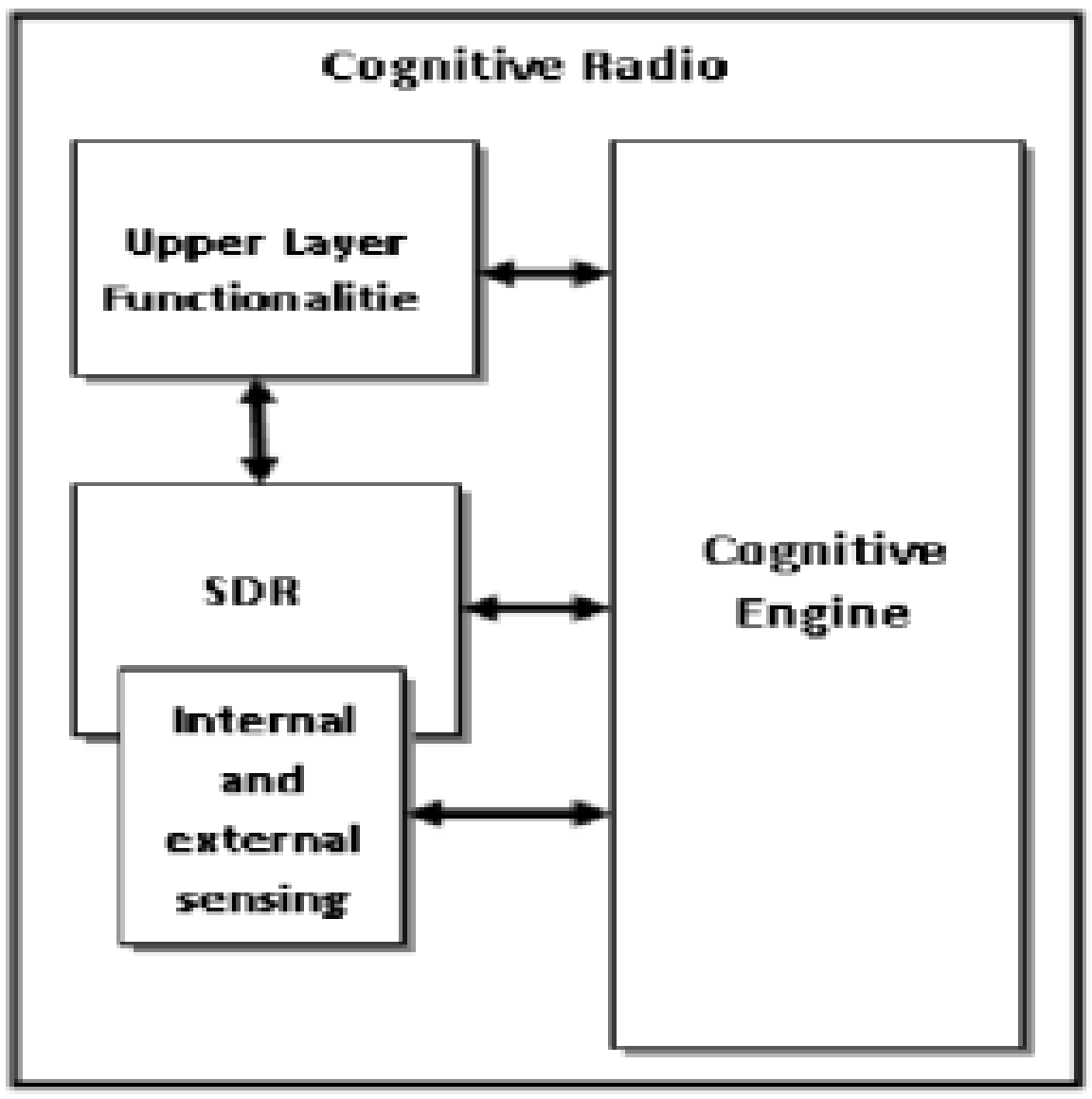

**Figure I.2:** Relation entre radio cognitive et radio logicielle [16]

La figure suivante décrit l'architecture d'un terminal radio cognitif AACR (Adaptative Aware Cognitive Radio) intégrant la radio logicielle SDR. Cette association montre qu'il est possible de combiner les avantages des techniques de transmission déjà existantes afin d'améliorer les capacités de la radio cognitive.

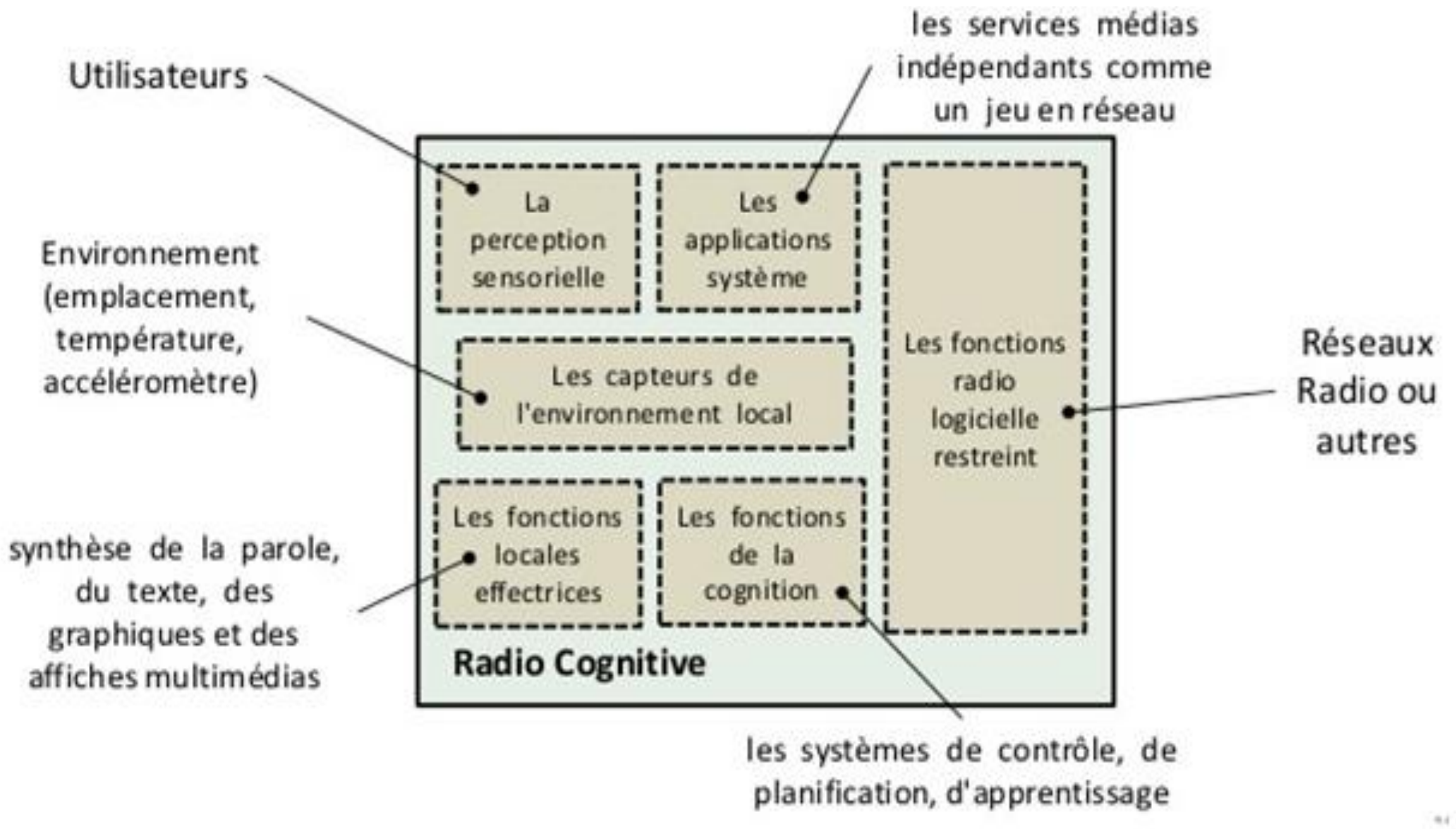

**Figure I.3:** Architecture du radio cognitive [17]

Les six composantes fonctionnelles CRA (Cognitive Radio Architecture) sont:

1. La perception sensorielle de l'utilisateur qui inclut l'interface haptique, acoustique, la vidéo et les fonctions de détection et de la perception. Les fonctions SP (sensory perception) de l'utilisateur peuvent inclure un matériel optimisé, par exemple, pour calculer les vecteurs de flux vidéo en temps réel pour aider la perception d'une scène.
2. Les capteurs de l'environnement local (emplacement, la température, l'accéléromètre, compas, etc.)
3. Les applications système (les services médias indépendants comme un jeu en réseau).
4. Les fonctions SDR qui incluent la détection RF (Radio Freqency ) et les applications radio de la SDR.
5. Les fonctions de la cognition (pour les systèmes de contrôle, de planification, de l'apprentissage).
6. Les fonctions locales effectrices (synthèse de la parole, du texte, des graphiques et des affiches multimédias).

## I.6.3 Principe de la radio cognitive

La radio cognitive est une technologie qui fait appel à l'intelligence des réseaux et des terminaux pour :

- Détecter les besoins de communications des utilisateurs en fonction de l'utilisation.
- Fournir des ressources radio et les services sans fil les plus appropriés à ces besoins.

Pour ce faire, la radio cognitive doit disposer de trois principales capacités [5] :

- Conscience **:** c'est la capacité à prendre conscience de son environnement. Un terminal radio cognitif associera donc environnement spatial, spectral et comportement des usagers pour une meilleure prise en conscience du provisionnement en ressources et un meilleur service.
- Adaptation : c'est la capacité à s'adapter soit à l'environnement (spectral ou technologique), soit à l'utilisateur (besoins et sécurité). L'adaptation à l'environnement spectrale est la capacité à choisir les meilleures bandes de fréquences et ainsi optimiser l'utilisation du spectre. Cela revient à :
  - Connaître l'occupation des bandes de fréquences en temps réel.
  - Adapter les puissances émises / marges en réception / Formes d'ondes.
  - Adapter le débit en temps réel en fonction de la place disponible.
  - Prendre en compte un partage temporel.

L'adaptation à l'utilisateur quant à elle se définit comme la capacité à reconnaître l'utilisateur. Cela est possible grâce à des techniques telles que la reconnaissance vocale et la biométrie.

- La cognition: La cognition regroupe les divers processus mentaux allant de l'analyse perceptive de l'environnement à la commande motrice (en passant par la mémorisation, le raisonnement, les émotions, le langage…). Cette définition dépasse donc le seul cadre de la cognition humaine ou animale. La radio cognitive est donc un système qui peut percevoir son environnement, l'analyser, le mémoriser et agir en conséquence.

## I.6.4 Cycle cognitif

Le cycle cognitif consiste en différentes étapes comme indiqué dans la figure I.6.
La figure décrite en détail ce cycle commençant par l'étape d'observation jusqu'à l'étape d'action afin de permettre à la radio cognitive d'interagir avec son environnement.
Les systèmes cognitifs observent, orientent, planifient, décident et agissent, tout en apprenant de leur environnement afin d'être plus efficaces au fil du temps. Les différentes étapes du cycle cognitif sont les suivantes :

- Observation : Extraire plusieurs informations à partir de l'environnement comme la fréquence radio, le type de données transmises (audio, vidéo, etc.), la position, etc.
- Orientation : Evaluer la situation et déterminer si elle est familière et réagir immédiatement, si nécessaire.
- Planification : Identifier les actions alternatives à prendre.
- Décision : Décider entre les actions candidates, en choisissant la meilleure d'entre elles.
- Action : Agir sur l'environnement en effectuant, par exemple, des modifications au niveau de la fréquence radio.
- Apprentissage automatique : S'informer à partir de l'expérience acquise à travers l'observation de l'environnement.

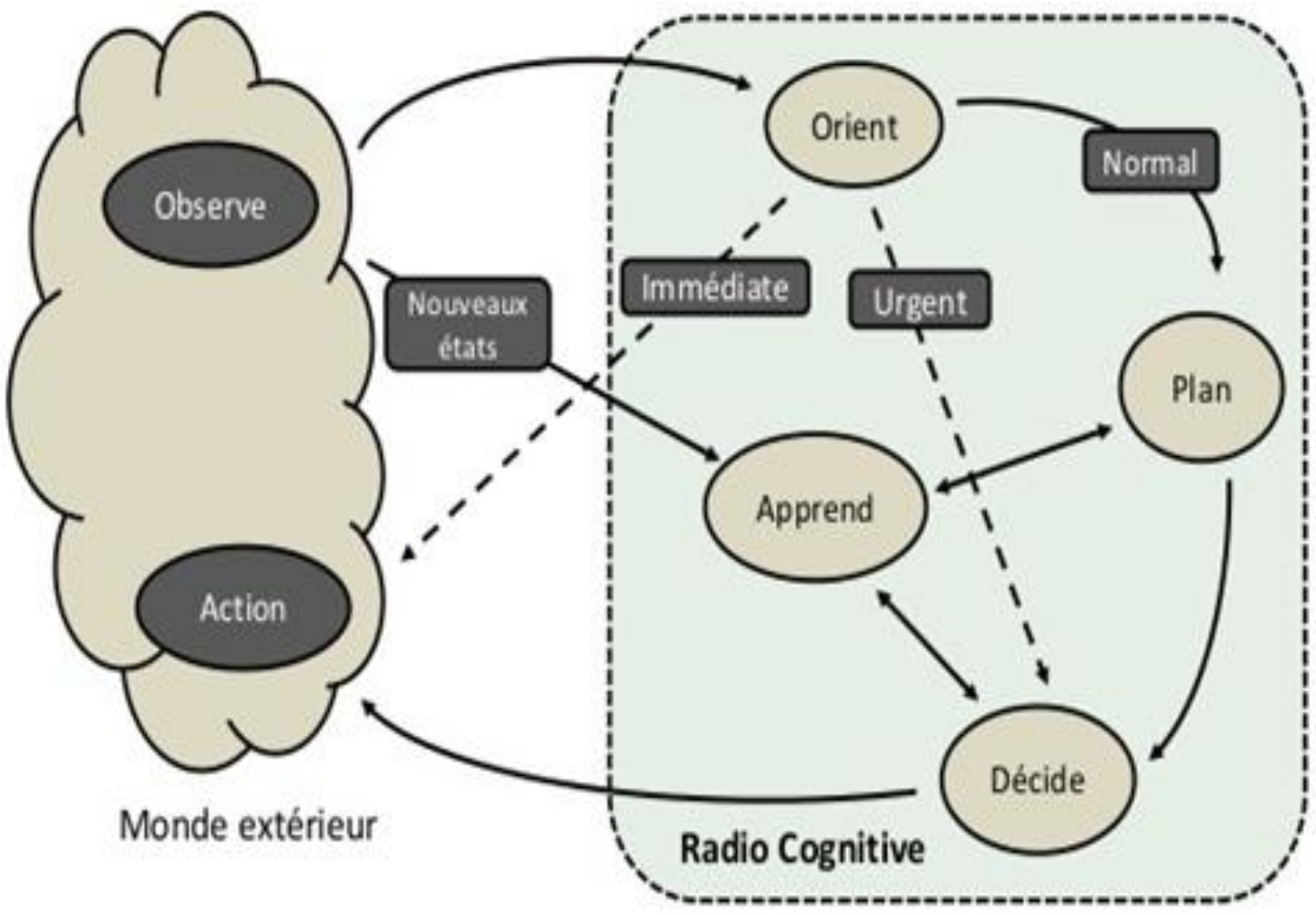

**Figure I.4:** Cycle de cognition de Mitola [18]

## I.6.5 Domaines d'application de la radio cognitive

Parmi les domaines d'application de la radio cognitive, on peut citer :

- **Réseaux de sécurité publique**

Les réseaux de sécurité publique peuvent exploiter le réseau de radio cognitive. Ils sont utilisés pour les communications entre la police, les pompiers et le personnel paramédical. De tels réseaux sont également contestés par la quantité limitée de spectre attribué, même avec les extensions récentes des bandes de fréquences de sécurités publiques allouées, le personnel de la sécurité publique ne dispose pas d'une technologie d'opérer dynamiquement à travers les différents segments du spectre. La radio cognitive peut offrir à ces réseaux, plus de bande passante par l'accès opportuniste au spectre. En outre, un RRC de la sécurité publique peut fournir une amélioration de la QoS permettant l'interopérabilité entre les différents services de sécurité publique [19].

- **Secours aux sinistrés et les réseaux d'urgence**

Les catastrophes naturelles comme les ouragans, les tremblements de terre, les incendies de forêt, ou autres phénomènes imprévisibles provoquent habituellement l'effondrement de l'infrastructure de communication. Il en résulte qu'un ensemble de réseaux coexistant partiellement ou totalement qui ont été précédemment déployés sont devenus déconnectés. Pendant ce temps, il y a un besoin urgent d'un moyen de communications pour aider les équipes de secours afin de faciliter l'aide organisée et la localisation des survivants de la catastrophe. Pour cela un RRC peut être utilisé pour une telle situation d'urgence [19].

- **Réseaux militaires**

Les communications sécurisées dans les champs de bataille modernes sont devenues une tâche très difficile sachant qu'un réseau de communication du champ de bataille est le seul moyen de communications entre les soldats, les véhicules armés et les autres unités dans le champ de bataille entre eux ainsi qu'avec le siège. Ceci implique que les réseaux ne nécessitent pas seulement une importante quantité de bande passante, mais aussi une garantie de communication sûre et fiable pour transporter des informations vitales. La radio cognitive est la technologie clé pour la réalisation de ces réseaux densément déployés en se basant sur des stratégies d'accès au spectre opportuniste [20].

- **Coexistence de différentes technologies sans fil**

La radio cognitive est une solution qui fournit la coexistence pour les nouvelles technologies sans fil qui sont en cours d'élaboration afin de réutiliser des fréquences radio allouées à d'autres services sans fil (service TV par exemple).

- **Services de cyber santé (eHealth services)**

Généralement tous les dispositifs médicaux de soins utilisés sont sans fil et sont sensibles aux EMI (interférences électromagnétique). D'autre part, différents dispositifs biomédicaux (équipement et appareil chirurgicaux, de diagnostic et de suivi) utilisent la transmission RF (Radio Frequency). Dans ce cas la radio cognitive peut être appliquée pour l'utilisation du spectre de ces dispositifs mais doit éviter toute interférence [21].

## I.7 Conclusion

Dans ce chapitre nous avons présenté la radio cognitive qui est une des solutions pour l'amélioration de la gestion et de l'utilisation du spectre radio. Nous avons présenté la gestion spectrale, l'architecture de la radio cognitive en intégrant la radio logicielle, ainsi que ses domaines d'applications. La radio cognitive est apparue comme une technologie clé pour permettre un accès opportuniste au spectre. L'objectif principal de cette gestion du spectre est d'obtenir un taux maximum d'exploration du spectre radio.

# Chapitre II : Les métaheuristiques

## II.1 Introduction

Les méthodes de résolution exactes consistent généralement à énumérer, souvent de manière implicite l'ensemble des solutions tel que son avantage est la certitude d'obtenir une solution optimale, mais le temps d'exécution est prohibitif. Ce pendant on peut trouver des solutions de bonne qualité sans garantie d'optimalité mais le temps de calcul est plus réduit. Pour cela on applique les métaheuristiques [22].

Les métaheuristiques constituent un ensemble de méthodes utilisées en intelligence artificielle et en recherche opérationnelle pour résoudre des problèmes d'optimisation difficiles. L'objectif des métaheuristiques est d'explorer l'espace de recherche efficacement afin de déterminer des solutions (presque) optimales.

Les techniques des algorithmes de ce type vont de la simple procédure de recherche locale à des processus d'apprentissage complexes [23]. Les métaheuristiques sont en général non-déterministes et ne donnent aucune garantie d'optimalité.

Dans ce chapitre, nous allons présenter les métaheuristiques en citant leurs deux grandes classes avec des exemples d'algorithmes. Nous donnons également une brève explication de l'algorithme SFLA qui est l'algorithme implémenté dans le cadre de ce travail.

## II.2 Terminologie

Le mot métaheuristique est composé de deux mots grecs:

- Heuristique qui vient du verbe heuriskein (ευρισκειν) et qui signifie « trouver/découvrir ».
- Méta qui est un suffixe signifiant « au-delà », dans un niveau supérieur.

## II.3 Heuristique et métaheuristique

### II.3.1 Heuristique

Une heuristique est un algorithme approché qui permet d'identifier en temps polynomial au moins une solution réalisable rapide, pas obligatoirement optimale. L'usage d'une heuristique est efficace pour calculer une solution approchée d'un problème et ainsi accélérer le processus de résolution exacte. Généralement une heuristique est conçue pour un problème particulier, en s'appuyant sur sa structure propre sans offrir aucune garantit quant à la qualité de la solution calculée [24].

### II.3.2 Métaheuristique

Les heuristiques ont rencontré des difficultés pour avoir une solution réalisable et de bonne qualité aux problèmes d'optimisation difficiles, pour cela les métaheuristiques ont fait leur apparition. Ces algorithmes sont plus complets et complexes qu'une simple heuristique, et permettent généralement d'obtenir une solution de très bonne qualité pour des problèmes issus des domaines de la recherche opérationnelle ou de l'ingénierie dont la résolution du problème nécessite un temps élevé ou une grande mémoire de stockage. Dans les métaheuristiques, le rapport entre le temps d'exécution et la qualité de la solution trouvée est très intéressant par rapport aux différents types d'approches de résolution [25]. Elles utilisent des processus aléatoires et itératifs pour rassembler de l'information, d'explorer l'espace de recherche et de résoudre les problèmes d'optimisation combinatoire. Une heuristique est utilisée à un problème donné. Par contre, une métaheuristique peut être adaptée pour différents types de problèmes. Plusieurs domaines sont appliqués par cette dernière tel que : la biologie (algorithmes évolutionnaires et génétiques), la physique (recuit simulé), et aussi l'éthologie (algorithmes de colonies de fourmis).

## II.4 Notions des métaheuristiques

Il existe plusieurs notions des métaheuristiques :

- Voisinage : la notion de voisinage est le principe général le plus utilisé pour la conception d'une métaheuristique. Cette notion structure l'espace de recherche dans le sens où elle permet de définir des sous-ensembles de solutions. En particulier, on peut établir plusieurs structures de voisinage à partir d'une solution donnée, selon la transformation que l'on s'autorise, celle-ci étant définie comme une application V: S→P(S). Chaque structure de voisinage fournit un ensemble de solutions, précis via la transformation définie.

Par exemple, V (S) est un sous-ensemble de configurations de S qui est atteint à partir d'une transformation donnée de S, et S' ∈ V (S) est une solution dite voisine de S [26].

- Diversification : permet de générer des diverses solutions afin d'explorer l'espace de recherche dans une échelle globale [27].
- Intensification : le processus d'intensification vise à forcer une solution donnée à tendre vers l'optimum local de la zone à laquelle elle est attachée [28].

Il est nécessaire de bien doser l'usage de ces deux notions (diversification, intensification) afin que l'exploration puisse rapidement identifier des régions de l'espace de recherche qui contiennent des solutions de bonne qualité, sans perdre trop de temps à exploiter des régions moins prometteuses.

- Mémoire et apprentissage : la mémoire est le support de l'apprentissage, qui permet à l'algorithme de ne tenir compte que des zones où l'optimum global est susceptible de se trouver, évitant ainsi les optima locaux.

La figure suivante représente les trois phases d'une métaheuristique itérative.

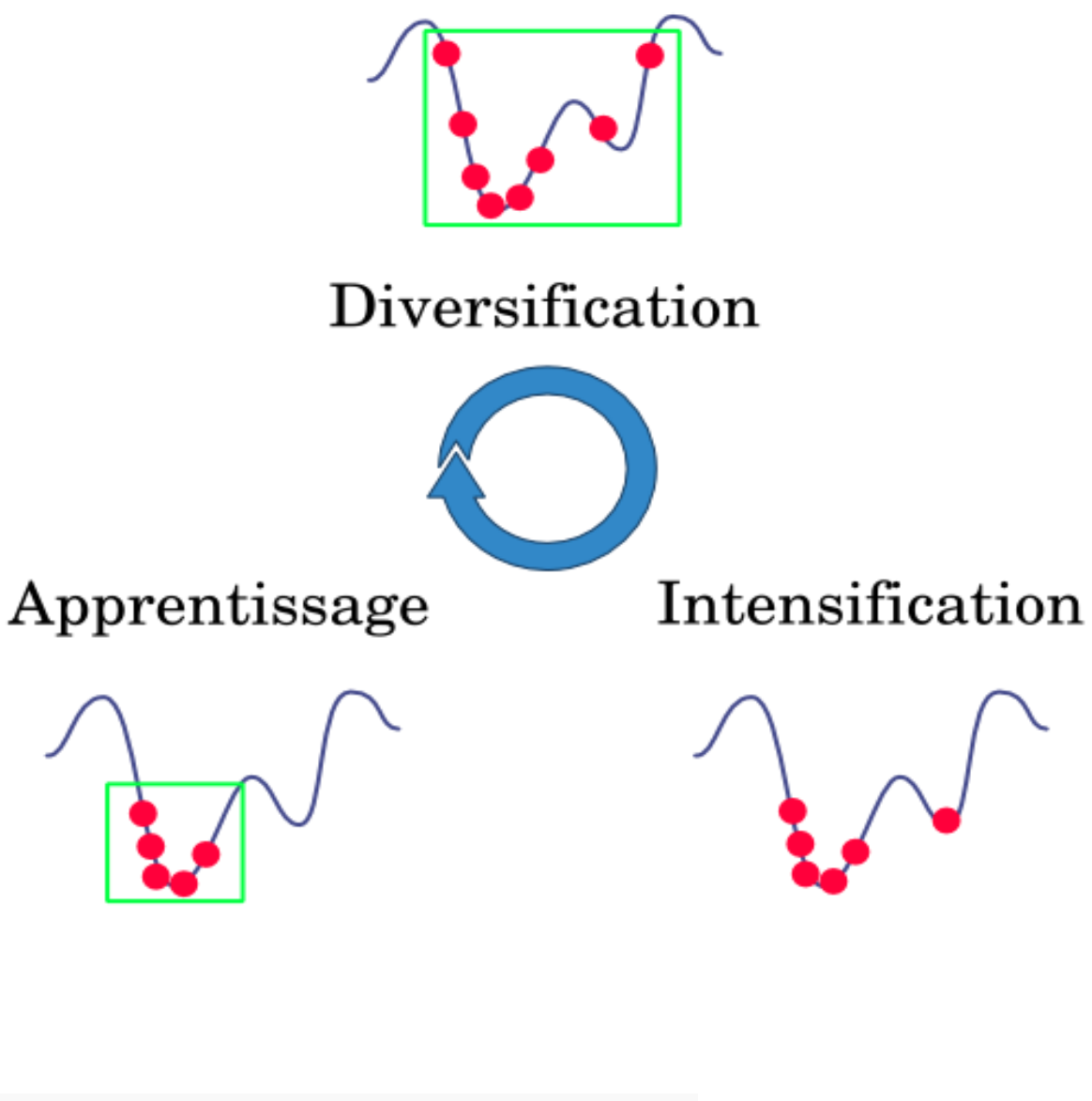

: représentent l'échantillonnage de la fonction objectif

**Figure II.1:**Les trois phases d'une métaheuristique itérative

## II.5 Classification des métaheuristiques

La figure suivante représente la classification des métaheuristiques.

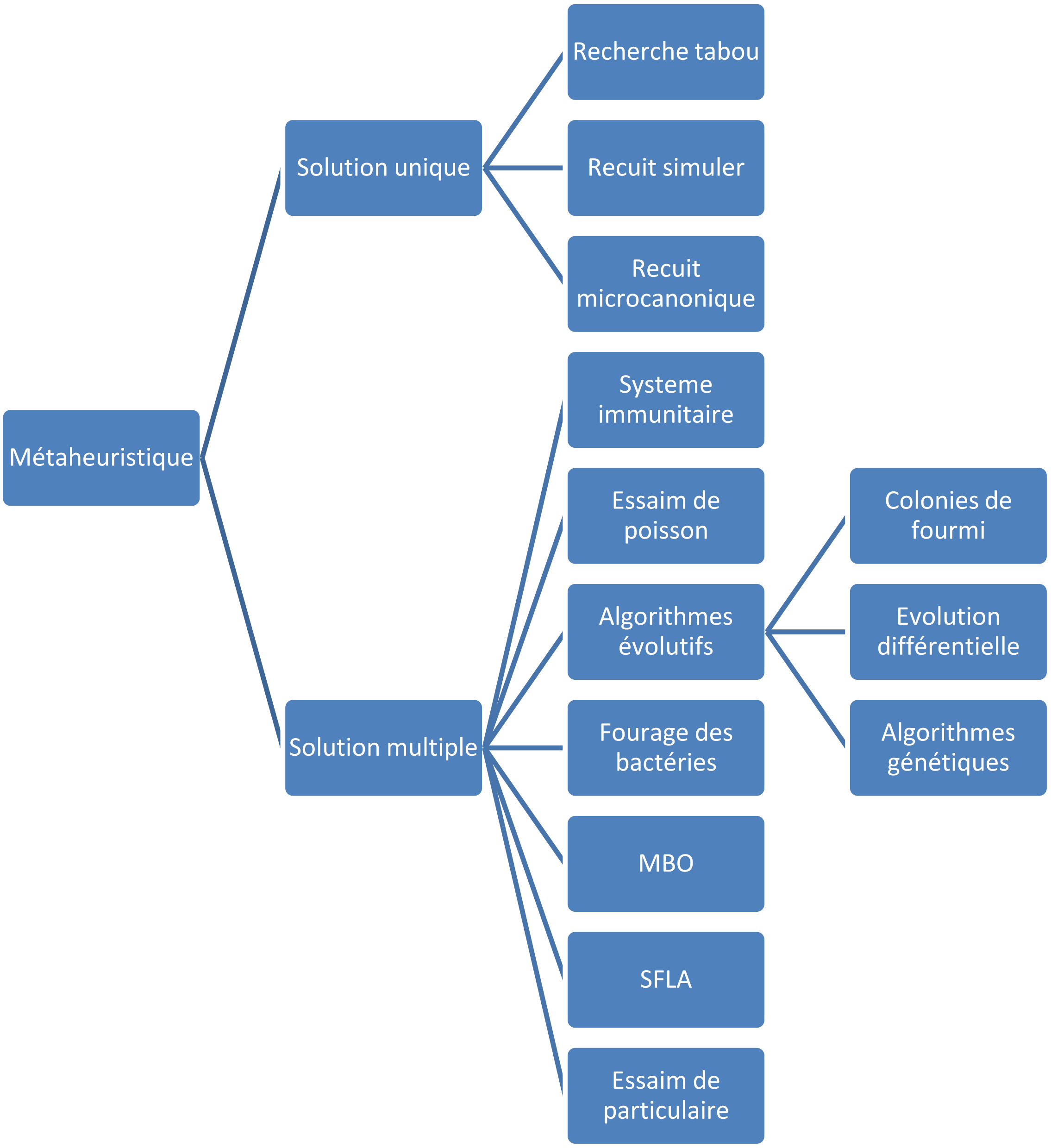

**Figure II.2:** Classification des métaheuristiques

Les métaheuristiques sont divisées en deux grandes classes :

- **Les métaheuristiques à solution unique :** sont appelées méthodes de recherche locale ou méthodes de trajectoire, commencent avec une seule solution initiale, et par une amélioration itérative, en construisant un ensemble de solutions de coûts décroissants pour un problème de minimisation. Le processus s'arrête si la solution courante ne peut pas être améliorée ou si le nombre d'itérations maximum est atteint. Quelques exemples typiques de cette classe : recuit simulé, recherche tabou, et le recuit microcanonique, etc [29].

- **Les métaheuristiques à solution multiple :** travaillent sur un ensemble de points de l'espace de recherche en commençant avec une population de solution initiale puis de l'améliorer au fur et à mesure des itérations.

Elles ont une capacité à parcourir les grands espaces de recherche mais n'ont pas un fort pouvoir d'intensification et peuvent avoir tendance à converger lentement. L'objectif de ces méthodes est d'utiliser la population comme un facteur de diversité. Cette métaheuristique englobe : les algorithmes évolutifs (Colonies de fourmi, algorithmes génétiques, Evolution différentielle), Essaim particulaire, SFLA, etc) [30].

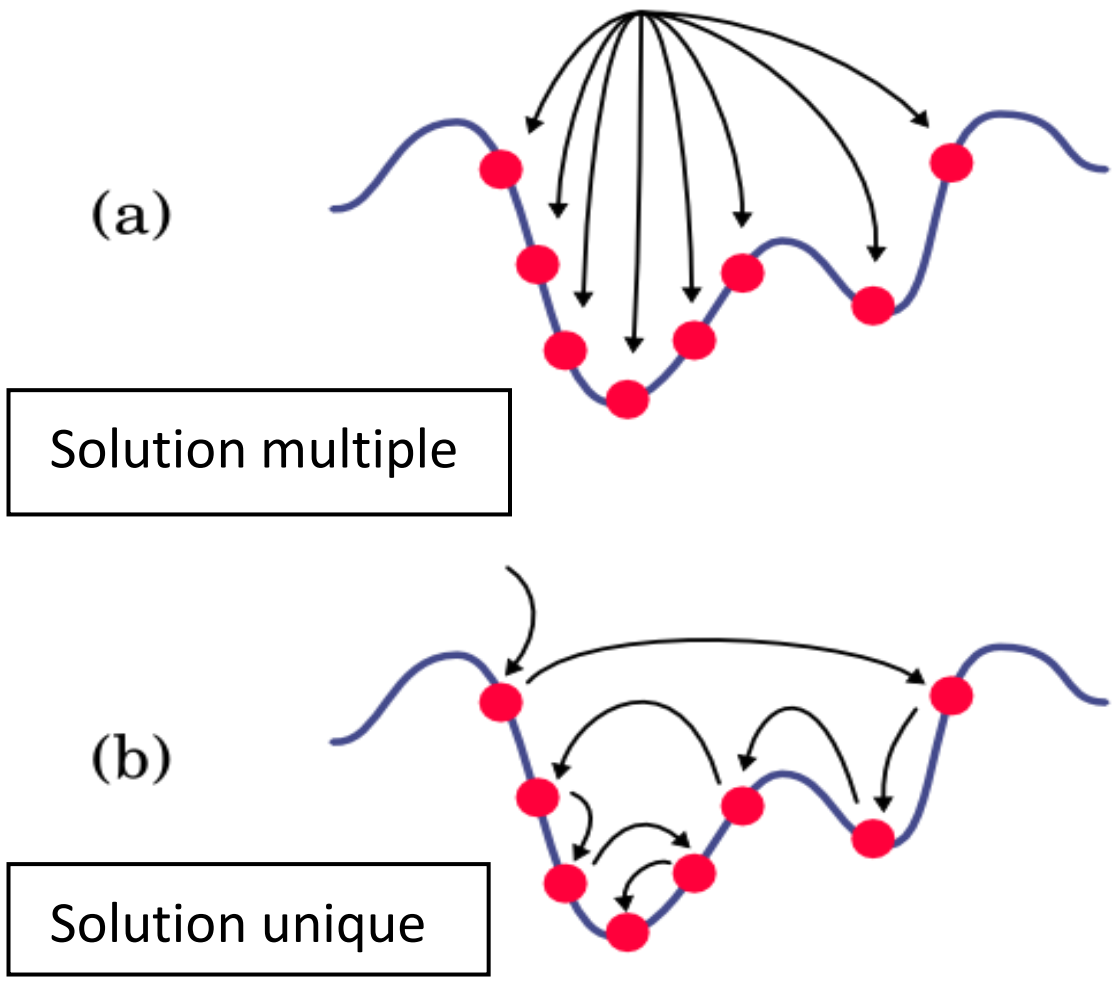

**Figure II.3:** Exemple d'une solution multiple et unique

# II.6 Algorithme SFLA

## II.6.1 Présentation

L'algorithme par saut de grenouilles « Shuffled Frog-Leaping Algorithm » (SFLA) a été développé pour résoudre des problèmes d'optimisation combinatoire. Le SFLA est un algorithme de recherche coopérative basé sur la population, il est inspiré par les systèmes naturels. Il se compose d'un ensemble de grenouilles (chacune représente une solution au problème) partitionnées en différent groupes. Les grenouilles peuvent communiquer entre elles et améliorer leurs solutions par contamination (passant l'information). L'algorithme contient des éléments de recherche locale effectuée dans chaque groupe avec un échange d'information globale. L'information entre les différentes communautés circule par l'intermédiaire d'un processus de saut. Dans chaque communauté, les grenouilles fournissant la meilleure solution Xb et la plus mauvaise Xw. La grenouille donnant la meilleure solution dans la population entière est notée par Xg. Pendant l'évolution d'une communauté, c.-à-d., pendant l'exploration locale, la plus mauvaise grenouille effectue un saut vers la meilleure Xb selon la règle suivante :

$$S = r \times (|Xb - Xw|) \quad (0 < r < 1)$$
$$Xw' = Xw + S \qquad S < Smax)$$

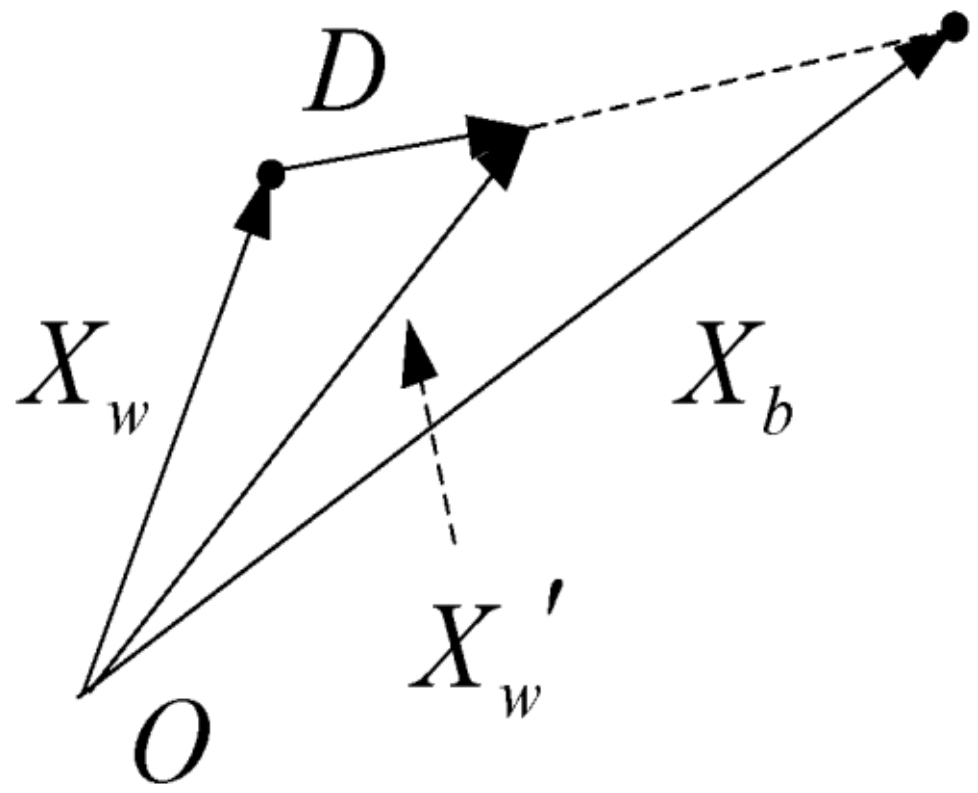

**Figure II.4:** Règle du saut de grenouille [31]

Afin d'assurer l'exploration globale, les communautés sont mélangées et réorganisées à nouveaux pour former une nouvelle population, ce mécanisme est répété jusqu'à satisfaire un critère d'arrêt [32].

## II.6.2 Principes de SFLA

Le SFLA se déroule sur 7 étapes comme suit :

**Etape 1 :** Fixer la taille *F* de la population, le nombre $m$ de communauté et le nombre *N* d'itérations.

**Etape 2 :** Générer aléatoirement une population de *F* solutions et évaluer chaque solution.

**Etape 3 :** Trier la population et déterminer la meilleure solution $X$g.

**Etape 4 :** Partitionner la population en $m$ communautés.

**Etape 5 :** Recherche locale, pour chaque communauté, répéter pour *N* itérations :

- Déterminer la meilleure solution $Xb$ et la mauvaise solution $Xw$.
- Calculer la solution $Xw$ *'* à partir de $Xb$.
- **Si** $Xw$ *'* est meilleure que $Xw$ **alors** remplacer $Xw$ par $Xw$ *'*.
- **Sinon** calculer $Xw$ *'* à partir de $X$g.
  - **Si** $Xw$ *'* est meilleure que $Xw$ **alors** remplacer $Xw$ par $Xw$ *'*.
  - **Sinon** générer aléatoirement $Xw$ *'* et remplacer $Xw$ par $Xw$ *'*.
  - **Finsi**
- **Finsi**

**Etape 6 :** Regrouper les m communautés pour former à nouveau la population.

**Etape 7 :** Aller à l'étape 3 si le critère d'arrêt n'est pas atteint.

## II.7 Conclusion

Nous avons présenté dans ce chapitre les métaheuristiques et leurs deux grandes classes ; les métaheuristique à solution unique et les métaheuristique à population de solution. Elles consistent à des méthodes approchées adaptables à une grande variété de problèmes d'optimisation combinatoire et mènent à des résultats pertinents. Certaines métaheuristiques présentent l'avantage d'être simples à mettre en œuvre, comme le cas du recuit simulé, d'autres sont bien adaptées à la résolution de certaines classes de problème, très contraints, comme le système de colonies de fourmis. La qualité des solutions trouvées par les métaheuristiques dépend de leur paramétrage, et de l'équilibre entre un balayage de tout l'espace des solutions (diversification) et une exploration locale (l'intensification) [25].

Parmi les méthodes proposées, nous nous intéresserons dans le chapitre suivant à l'algorithme SFLA qui est un outil efficace pour résoudre des problèmes d'optimisation combinatoire.

# Chapitre III: Evaluation et test

## III.1 Introduction

Dans de nombreux problèmes scientifiques et techniques, il est important de trouver le minimum ou le maximum d'une fonction de nombreuses variables. Nous avons traité dans ce contexte le problème de l'optimisation de la qualité de service dans un réseau de radio cognitive en utilisant l'algorithme de grenouilles sautant SFLA introduit dans le chapitre précédent. Cet algorithme est à l'origine développé par M. Eusuff et K. Lansey en 2003 [33], elle a été conçue comme une métaheuristique pour effectuer une recherche de la solution d'un problème d'optimisation combinatoire. L'objectif de ce chapitre est de montrer l'intérêt de l'application du SFLA dans le contexte d'un réseau de radio cognitive.

## III.2 La fonction objectif

Les fonctions objectifs sont utilisées en optimisation mathématique pour guider un système à un état optimal et refléter la qualité d'une solution au problème. Cette solution, soit elle donne la valeur la plus élevée (maximum) ou la valeur la moins élevé (minimum).

Dans notre cas, ces fonctions objectifs peuvent améliorer la performance et répondre aux exigences de qualité de service prenant en considération les critères suivants :

1. Maximisation du débit.
2. Minimisation du taux d'erreur.
3. Minimisation de la consommation d'énergie.

La représentation mathématique de ces critères [34] est définit dans le tableau ci dessous :

| Critère | Fonction |
|---|---|
| **Maximisation du débit** | $\frac{\log_2(M)}{\log_2 M_{max}}$ |
| **Minimisation du taux d'erreur** | $1 - \frac{\log_{10}(0.5)}{\log_{10}(\overline{P_{be}})}$ |
| **Minimisation de la consommation d'énergie** | $1 - \frac{P_i}{n * P_{max}}$ |

**Tableau III.1:** Présentation mathématique des trois critères

## III.3 Définition des paramètres

Comme nous avons mentionné précédemment, l'émetteurs-récepteurs radio sont capables de surveiller les transmissions des PU (utilisateur primaire) et l'environnement radio, mais aussi de changer les paramètres de leurs transmissions, telles que la puissance de transmission, le type de modulation et l'atténuation du signal pour réussir ou améliorer l'efficacité de la communication.
Dans le cas des applications qui ne supportent pas un taux d'erreur élevé, la RC doit sélectionner le type de modulation qui garantit le taux d'erreur le plus bas.

Le tableau suivant représente le taux d'erreur en fonction du type de modulation.

| Types de modulation | Taux d'erreur (Pbe) |
|---|---|
| **B PSk** | $Q\left(\sqrt{\frac{P}{N}}\right)$ |
| **M-ary PSk** | $\frac{2}{\log_2(M)} Q\left(\sqrt{2 * \log_2(M) * \frac{P_i}{N}} * \sin\frac{\pi}{M}\right)$ |
| **M-ary QAM** | $\frac{4}{\log_2(M)}\left(1 - \frac{1}{\sqrt{M}}\right) Q\left(\sqrt{\frac{3 * \log_2(M)}{M-1}\ \frac{P_i}{N}}\right)$ |

**Tableau III. 2:** Présentation mathématique du taux d'erreur selon le type de modulation

Notons que les trois fonctions [35] citées dans le tableau III.2 utilisent la fonction d'erreur de Gauss Q(x) :

$$Q(x) = \frac{e^{-\frac{x^2}{2}}}{1.64x + \sqrt{0.7x^2 + 4}}$$

Les différentes variables utilisées ci-dessus sont définis comme suit :

| **Variable** | **Définition** |
|---|---|
| N | Le nombre de sous porteuses. |
| $P_i$ | La puissance du signal sur la sous porteuse i. |
| $P_{max}$ | La puissance maximale qu'on peut transmettre sur une seule sous-porteuse. |
| M | L'index de modulation. |
| $M_{max}$ | L'index de modulation maximal. |
| $\overline{P_{be}}$ | Le taux d'erreur moyen sur chaque sous porteuses. |
| N | Le taux d'atténuation. |

**Tableau III. 3:**Définition des différents paramètres

La relation de dépendance entre l'ensemble des paramètres dans chaque critère est déterminée comme suit :

| **Critère** | **Paramètre associés** |
|---|---|
| Maximisation du débit | M ↗ |
| Minimisation du taux d'erreur | P ↗, N ↘, M ↘ |
| Minimisation de la consommation d'énergie | P ↘ |

**Tableau III. 4:** Relation entre les paramètres dans les différents critères

Maximiser le débit traite le débit de données du système tel qu'il devrait être augmenté. Cet objectif basé principalement sur l'index de modulation qui doit être plus élevé (M↗).

Minimiser le taux d'erreur est un objectif de communication extrêmement commun. Cet objectif consiste à réduire au minimum le nombre d'erreurs par rapport à la quantité de bits transmis. La détermination du taux d'erreur théorique nécessite la transmission de nombreuses paramètres qui comprennent une augmentation de puissance d'émission (P ↗) et une diminution de l'index de modulation et d'atténuation du signal (M ↘, N ↘).

L'objectif de consommation d'énergie est utilisé pour diriger le système à un état d'utilisation d'énergie minimale. La chose la plus essentielle pour cette fonction objectif est de diminuer la quantité de la puissance de transmission (P ↘).

Ces trois fonctions objectifs doivent également contenir un rang quantifiable traduisant l'importance de chacun selon l'exigence de l'utilisateur. Plusieurs approches existent pour déterminer cette préférence. Nous avons utilisé une approche de somme pondérée qui nous permet de développer une fonction objectif unique pour chaque objectif et les cumuler pour créer une fonction objectif multiple.
Cette approche limite la somme des poids associé à 1 pour une pondération plus intuitive et non ambigu.

$$\text{Fonction Objectif} = (\text{poids}_1 * \text{fct}_{\text{maxDebit}}) + (\text{poids}_2 * \text{fct}_{\text{minErreur}}) + (\text{poids}_3 * \text{fct}_{\text{MinEnergie}})$$

Avec : $\sum_{i=1}^{3} poids_i = 1$ tell que $poids_i \epsilon [0,1]$

## III.4 Modes de transmission

La notion de QoS dans la radio cognitive ne consiste pas à tout optimiser à cause de conflit entre les différents objectifs. Ainsi, un mode de transmission peut être vu comme une association de poids aux différents objectifs de la radio, de sorte que la fonction la plus importante reçoive le poids le plus fort, la fonction conflictuelle reçoive le poids le plus faible et la troisième fonction reçoive le poids intermédiaire.

Dans le cadre de notre travail, nous adopterons trois modes de transmission:

### a. Mode urgence

Ce mode se fonctionne lorsqu'il est détecté que le type d'application gérée par l'utilisateur est sensible aux erreurs.

$$\text{Fonction Objectif} = (0.05 * \text{fct}_{\text{maxDebit}}) + (0.80 * \text{fct}_{\text{minErreur}}) + (0.15 * \text{fct}_{\text{MinEnergie}})$$

**b. Mode multimédia**

Il est utilisé quand l'utilisateur gère une application multimédia par exemple une vidéoconférence.

$$\text{Fonction Objectif} = (0.80 * fct_{maxDebit}) + (0.05 * fct_{minErreur}) + (0.15 * fct_{MinEnergie})$$

**c. Mode batterie faible**

Ce mode de fonctionnement est utilisé lors de la détection de faiblesse de la charge de batterie.

$$\text{Fonction Objectif} = (0.05 * fct_{maxDebit}) + (0.15 * fct_{minErreur}) + (0.80 * fct_{MinEnergie})$$

| Fonction objectif / Mode de transmission | Poids associé | | |
|---|---|---|---|
| | **Maximiser le débit** | **Minimiser le taux d'erreur** | **Minimiser la consommation d'énergie** |
| Urgence | 0.05 | 0.08 | 0.15 |
| Multimédia | 0.08 | 0.05 | 0.15 |
| Batterie faible | 0.05 | 0.15 | 0.08 |

**Tableau III. 5:** Poids associé au différent mode

L'obtention des paramètres de transmission se fait à l'aide d'une évaluation de la fonction objectif qui se base sur les paramètres de l'environnement et le choix de type de mode de transmission comme il est montré dans la figure suivante :

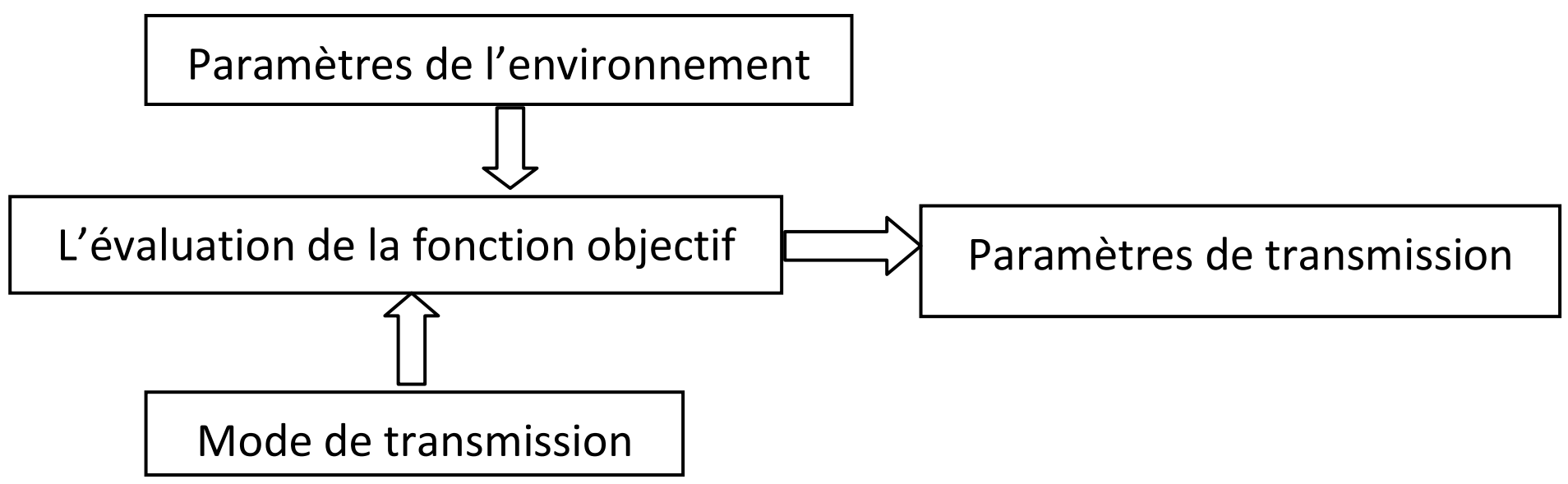

**Figure III.1:** Processus de prise de décision

## III.5 Présentation d'un canal

La transmission à travers le canal a besoin de trois paramètres : la puissance, le type de modulation et l'atténuation.

### a. Puissance

Une radio cognitive doit être capable de modifier sa puissance de transmission vu que cette dernière est spécifique à chaque bande de fréquence et ne doit en aucun cas dépasser un certain seuil même pour l'utilisateur primaire.

La plage des puissances de transmission prise varie entre 0,1 mW et 2,4808 mW avec un pas de 0,025 mW ce qui donne 94 valeurs de puissance. La valeur maximale de puissance est de 2,4808 mW a été choisie par apport à la valeur de la puissance maximale autorisée pour la bande U-NII (Unlicenced-National Information Infrastructure) : 5.15 GHz - 5.25 GHz est fixé à 2.5 mW [36].

### b. Modulation

Parmi les paramètres qu'une radio cognitive doit être capable de reconfigurer dynamiquement, c'est la modulation vu que la bande passante nécessaire pour transporter un signal dépend du type de modulation employée. Dans notre simulation, nous avons utilisé deux type de une modulation : la modulation QAM (Quadrature Amplitude Modulation) et la modulation PSK (Phase Shift Keying), avec un index de modulation (nombre de bits par symbole) qui varie entre 0 et 10 ($\mathbf{2^{\mathit{i}}}$ , $\mathbf{i \in [0, 10]}$ ).

### c. Atténuation

Lors de la propagation d'un signal, il s'atténue. Cela se traduit par la diminution relative de la puissance d'un signal au cours de sa transmission.

Dans le cas d'un environnement multicanaux dynamique, nous avons attribué pour chaque canal une valeur aléatoire compris dans l'intervalle $[\mathbf{0}\ ,\mathbf{1}]\ \mathbf{dB}$.

La figure suivante représente la structure d'un canal.

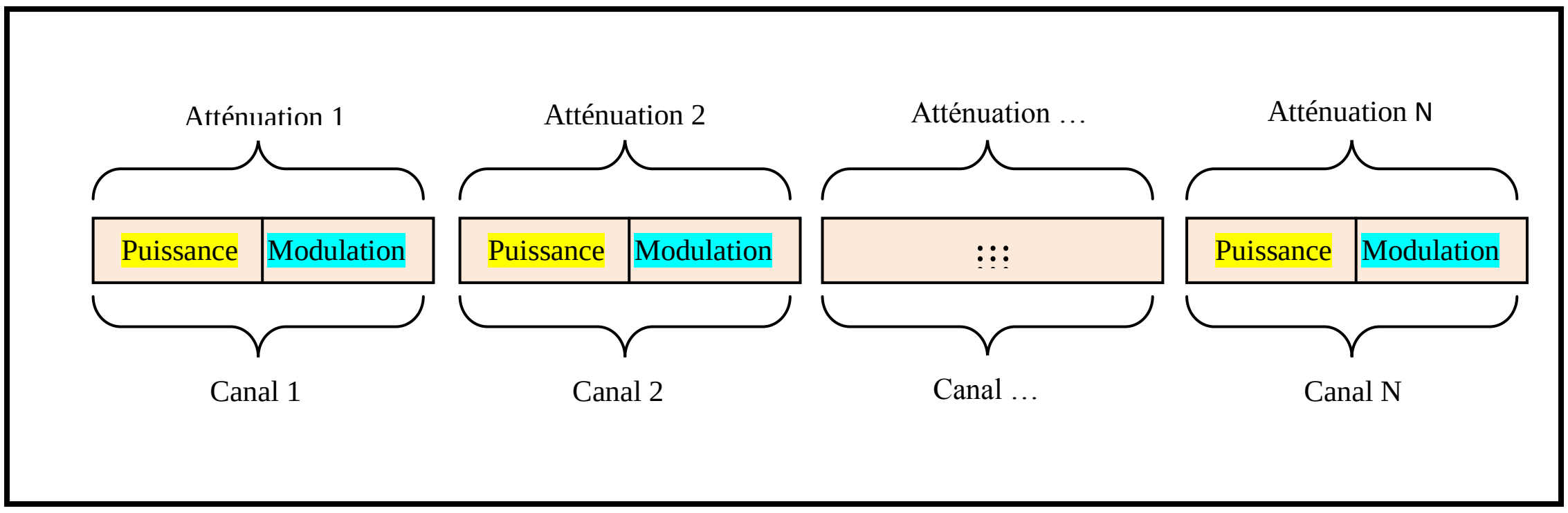

**Figure III.2:** Structure du canal

## III.6 Codage des paramètres

### a. Codage de puissance

| Codage | **0** | **2** | **3** | **4** | **…** | **92** | **93** |
|---|---|---|---|---|---|---|---|
| Valeur | 0.1 | 0.1256 | 0.151 | 0.176 | … | 2.429 | 2.488 mw |

**Tableau III.6:** Codage de puissance

### b. Codage de modulation

| Codage | 1 | 2 | 3 | 4 | 5 | 6 |
|---|---|---|---|---|---|---|
| Modulation | PSK | 2QAM | 4QAM | 8QAM | 16QAM | 32QAM |

| 7 | 8 | 9 | 10 | 11 |
|---|---|---|---|---|
| 64QAM | 128QAM | 256QAM | 512QAM | 1024QAM |

**Tableau III.7:** Codage de modulation

## III.7 Implémentation de l'algorithme SFLA

### a. Initialisation de la population

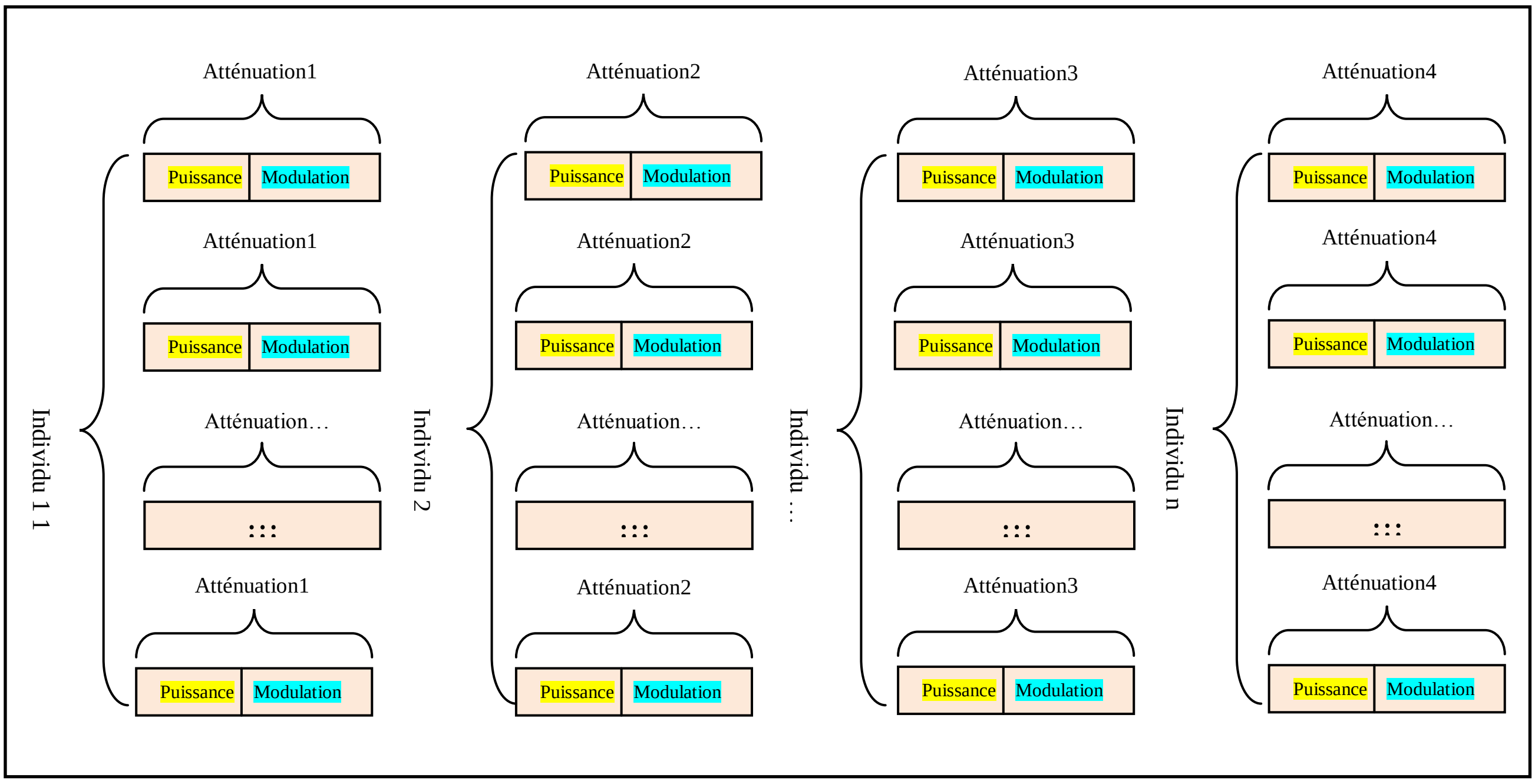

**Figure III.3:** Structure d'une population

### b. Amélioration des solutions

La population est triée par ordre décroissant selon la fonction objectif (fitness) et distribuée en plusieurs sous-ensembles différents décrits comme memeplexes. Pour améliorer les solutions, une recherche locale indépendante est réalisée pour chaque communauté.

Pendant l'amélioration d'une communauté, une nouvelle solution $X_{w'}$ sera calculée à partir de la meilleur solution local $X_b$ selon la règle du saut de grenouille. Si l'individu créé est meilleur que la mauvaise solution $X_w$, alors il la remplace.

Sinon, on applique la même règle en remplaçant cette fois ci $X_b$ par la solution globale $X_g$. Si la nouvelle solution $X_{w'}$ reste moins bonne que $X_w$, alors générer aléatoirement une autre solution qui remplace $X_w$.

La figure suivante montre le principe pour générer une nouvelle solution Xw'.

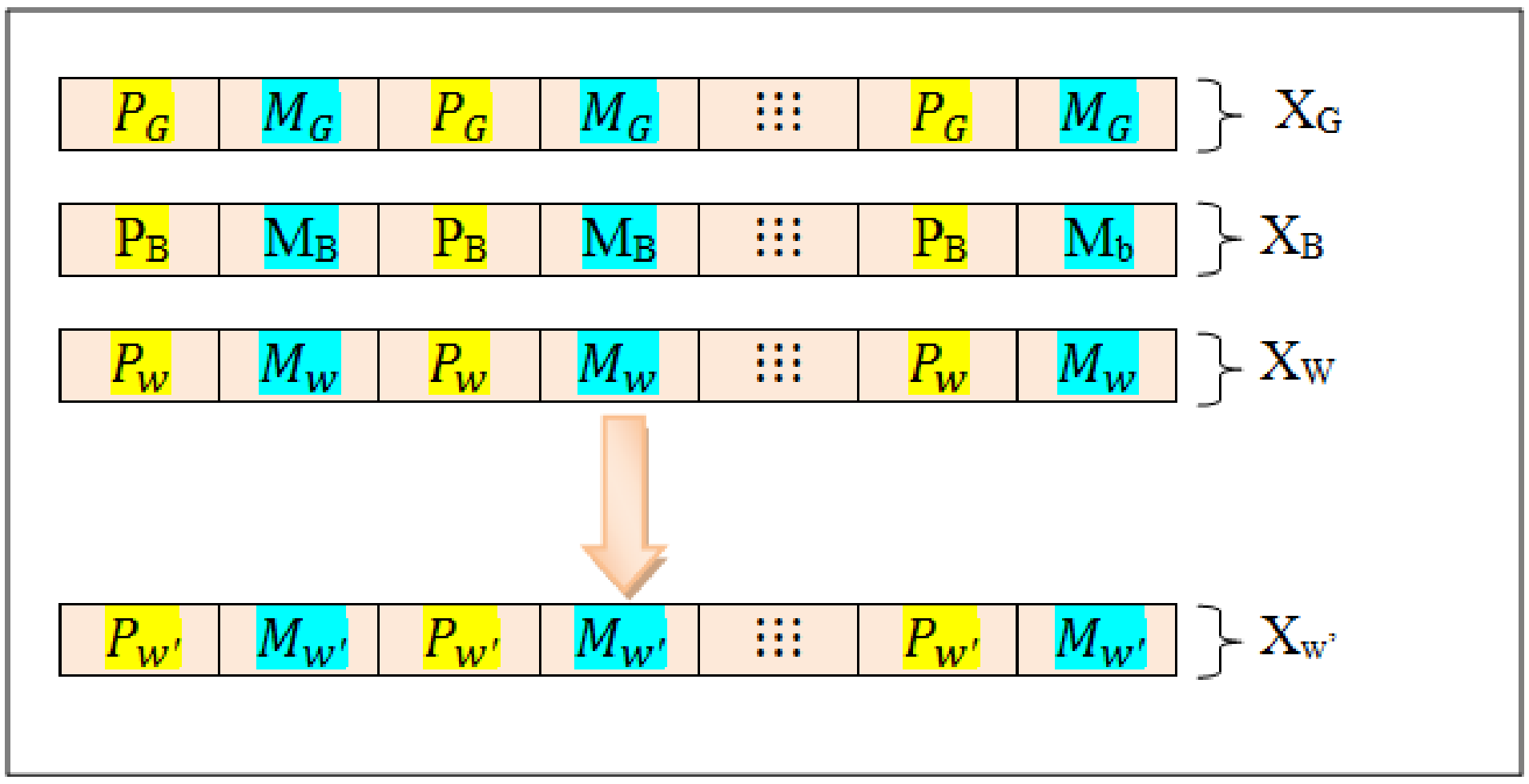

**Figure III.4:** Amélioration de la solution

1 er cas : Calculer $X_{w'}$ à partir de $X_B$

$Saut_{Puis} = r_x * (|P_B - P_W|)$
, $P_{w'} = P_w + Saut_{Puis}$

$Saut_{Mod} = r_x * (|M_B - M_W|)$
, $M_{w'} = M_w + Saut_{Mod}$

2eme cas : Calculer $X_{W'}$ à partir de $X_G$

$Saut_{Puis} = r_x * (|P_G - P_W|)$
, $P_{w'} = P_w + Saut_{Puis}$

$Saut_{Mod} = r_x * (|M_G - M_W|)$
, $M_{w'} = M_w + Saut_{Mod}$

3eme cas : Génération aléatoire
$P_{w'} = P_{aléa}$
$M_{w'} = M_{aléa}$

### c. Critère d’arrêt

Le test d’arrêt joue un rôle très important dans le jugement de la qualité des individus. Dans ce travail, le critère d’arrêt est définit par un nombre fixé a priori de générations.

## III.8 Expérimentation

Dans le genre de l'algorithme de grenouille sautant (SFLA), il est impossible de se baser uniquement sur une seule mesure pour pouvoir établir la meilleure solution au problème. Il est indispensable, dans ce cas de faire plusieurs simulations pour pouvoir confirmer un résultat.

### a. Convergence de la fonction objectif

Dans ce qui suit nous allons présenter le graphe de convergence de fitness en différents mode avec les paramètres suivants : 100 individus, 8 sous porteuses et 2000 générations.

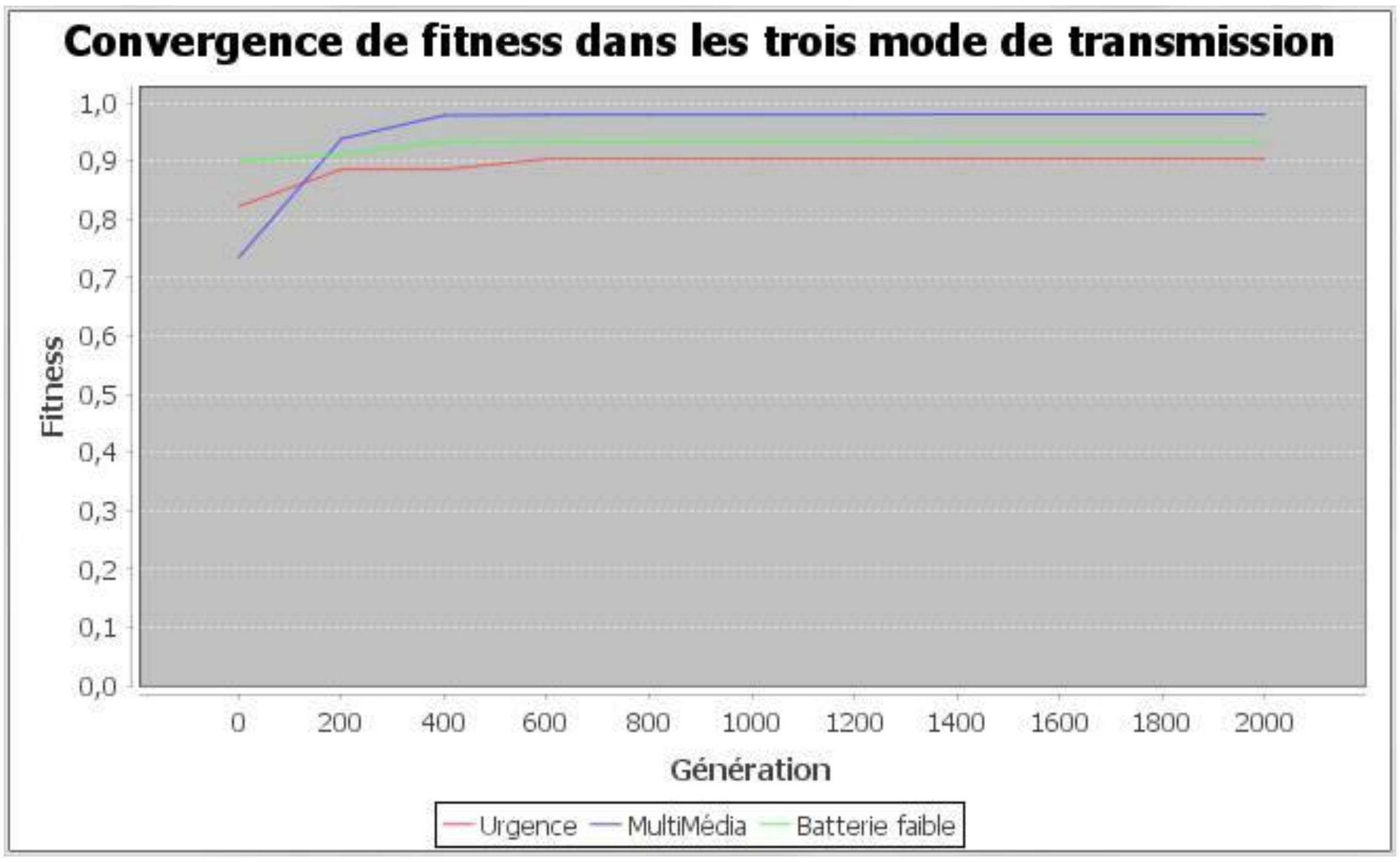

**Figure III.5:** Convergence de la fonction objectif dans les trois modes

Le tableau suivant résume les résultats du graphe précédent.

| Mode \ Fitness | Première génération | Dernière génération | Temps d'exécution (ms) |
|---|---|---|---|
| Urgence | 0.82260 | 0.90351 | 4938 |
| Multimédia | 0.73544 | 0.98029 | 4463 |
| Batterie faible | 0.90042 | 0.93434 | 4637 |

**Tableau III.8:** Fitness de la première et de la dernière génération dans les trois modes

D'après les résultats obtenus, nous remarquons que dans chaque mode, la fonction objectif est en amélioration depuis la première génération jusqu'à la dernière génération. En effet, le SFLA est capable d'atteindre une meilleure solution avec une convergence rapide. Toute fois, le mode multimédia est meilleur que les autres modes concernant la vitesse de convergence et la valeur de la meilleure solution.

### b. Fonction objectif en fonction du nombre de sous porteuses

L'étude présentée postérieurement présente les valeurs obtenues de la fonction objectif pour la meilleure solution dans chaque mode en fonction du nombre de sous porteuses avec une population de 50 individus et 10000 itérations.

| Mode / Nombre de canaux | Urgence | Multimédia | Batterie faible |
|---|---|---|---|
| 8 | 0,88181 | 0,98065 | 0,93186 |
| 16 | 0,84844 | 0,98208 | 0,93951 |
| 32 | 0,82832 | 0,98371 | 0,94522 |
| 64 | 0,81333 | 0,98436 | 0,95004 |
| 128 | 0,79951 | 0,98478 | 0,95242 |
| 256 | 0,79394 | 0,98489 | 0,93186 |
| 512 | 0,78943 | 0,98502 | 0,95421 |

**Tableau III.9:** Fitness en fonction du nombre de sous porteuses

La figure suivante représente les résultats du tableau précédent sous forme d'un histogramme.

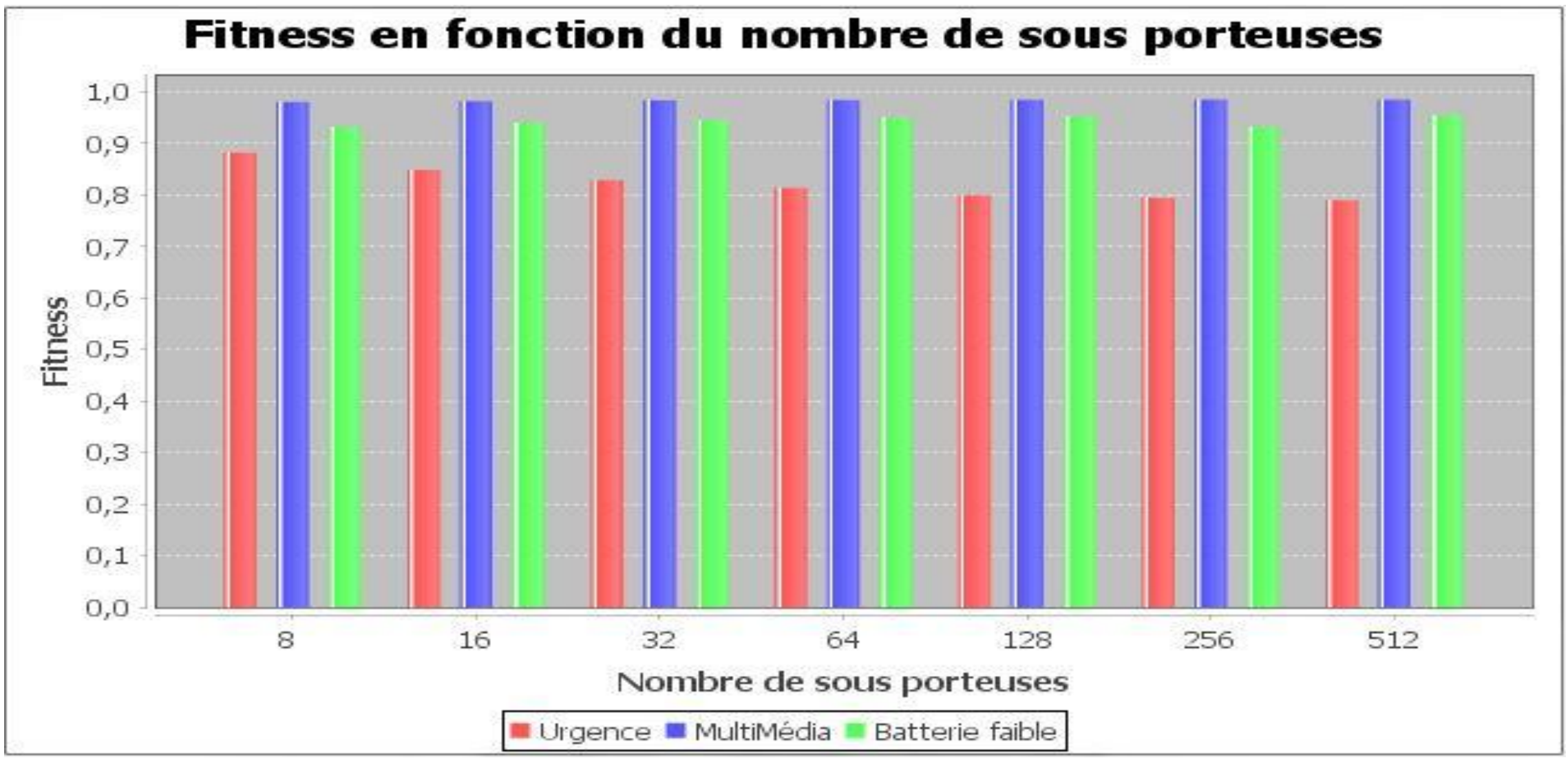

**Figure III.6:** Fitness en fonction du nombre de sous porteuses

A travers ces résultats obtenus, nous pouvons remarquer que quel que soit le nombre de sous porteuses, les valeurs de la fonction objectif sont intéressantes. Par exemple dans le mode multimédia la meilleure solution est dans les alentours de 0.98, en mode urgence elle est dans l'intervalle [0.78, 0.88] et elle est comprise entre 0.93 et 0.95 dans le mode batterie faible.

**c. Meilleur score de fitness dans les différents modes en fonction de nombre de groupe et de génération**

Cette expérimentation examine la fonction objectif où nous avons fixé la taille de la population à 100 et le nombre de sous porteuse à 8.

| | Génération / Groupe | 500 | 1000 | 1500 | 2000 |
|---|---|---|---|---|---|
| **Mode urgence** | **5** | 0.88195 | 0.88766 | 0.88793 | 0.89899 |
| | **10** | 0.88424 | 0.88801 | 0.89717 | 0.88975 |
| | **15** | 0.89128 | 0.90162 | 0.90607 | 0.90518 |
| | | | | | |
| **Mode multimédia** | **5** | 0.97993 | 0.98078 | 0.98007 | 0.98102 |
| | **10** | 0.98050 | 0.98079 | 0.98121 | 0.98153 |
| | **15** | 0.98041 | 0.98106 | 0.98133 | 0.98211 |
| | | | | | |
| **Mode batterie faible** | **5** | 0.93229 | 0.93179 | 0.93359 | 0.93457 |
| | **10** | 0.93727 | 0.93336 | 0.93788 | 0.93620 |
| | **15** | 0.93970 | 0.93603 | 0.93204 | 0.93695 |

**Tableau III.10:** Fitness dans les différents modes selon le nombre de groupe et de génération

D'après le tableau III.10, on constate que la fonction objectif de la meilleure solution s'augmente en augmentant le nombre de groupe et le nombre de génération ce qui donne aux individus plus de chance pour s'améliorer.

On observe également que le mode multimédia reste toujours le plus adaptable par rapport aux autres modes de communications.

Nous pouvons conclure que le SFLA a été en mesure pour améliorer la qualité de service et de répondre aux besoins de la radio cognitive.

Les figures III.7, III.8 et III.9 résument les résultats obtenus dans le tableau précédent.

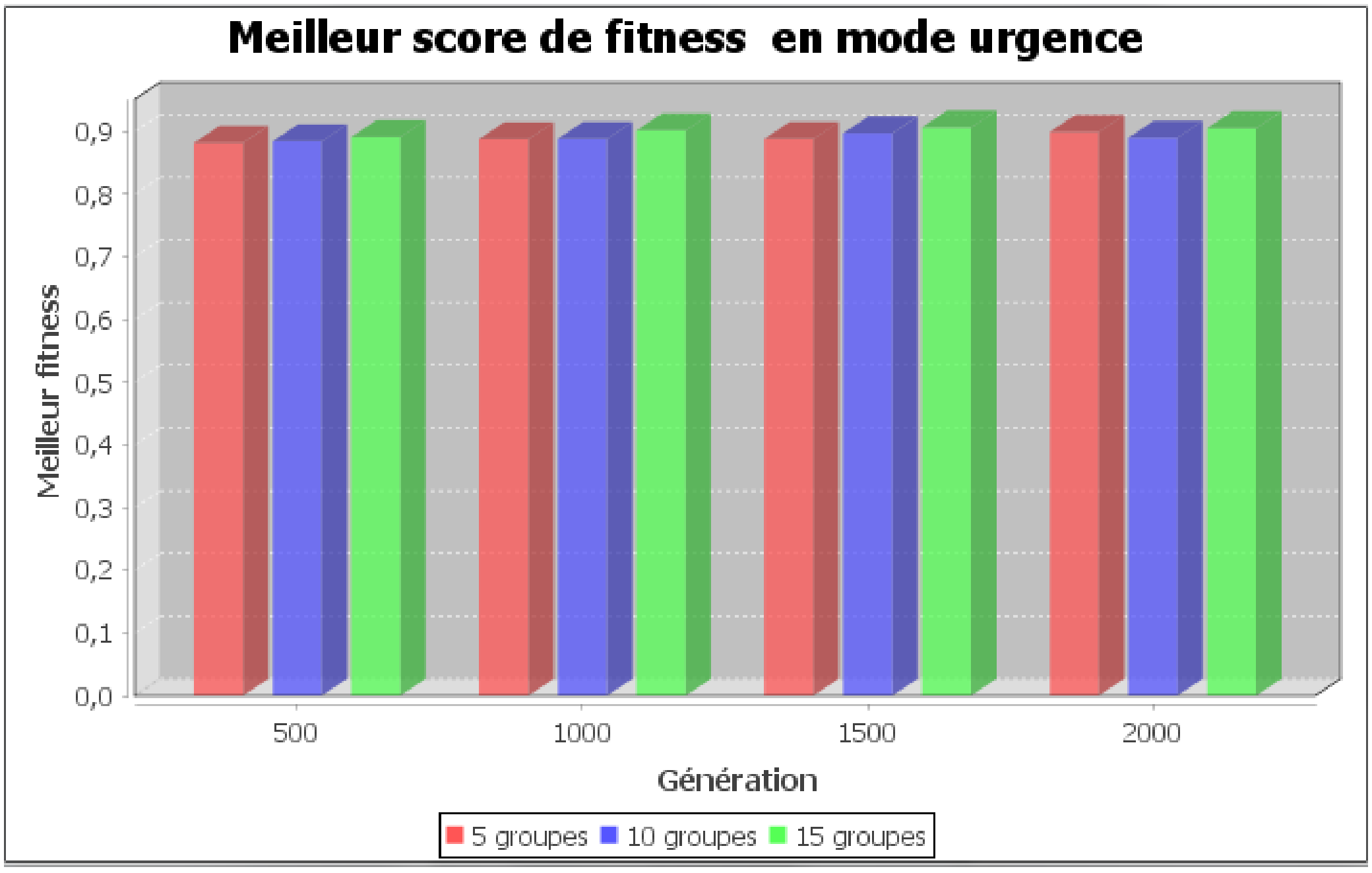

**Figure III.7:** Meilleure score de fitness en mode urgence

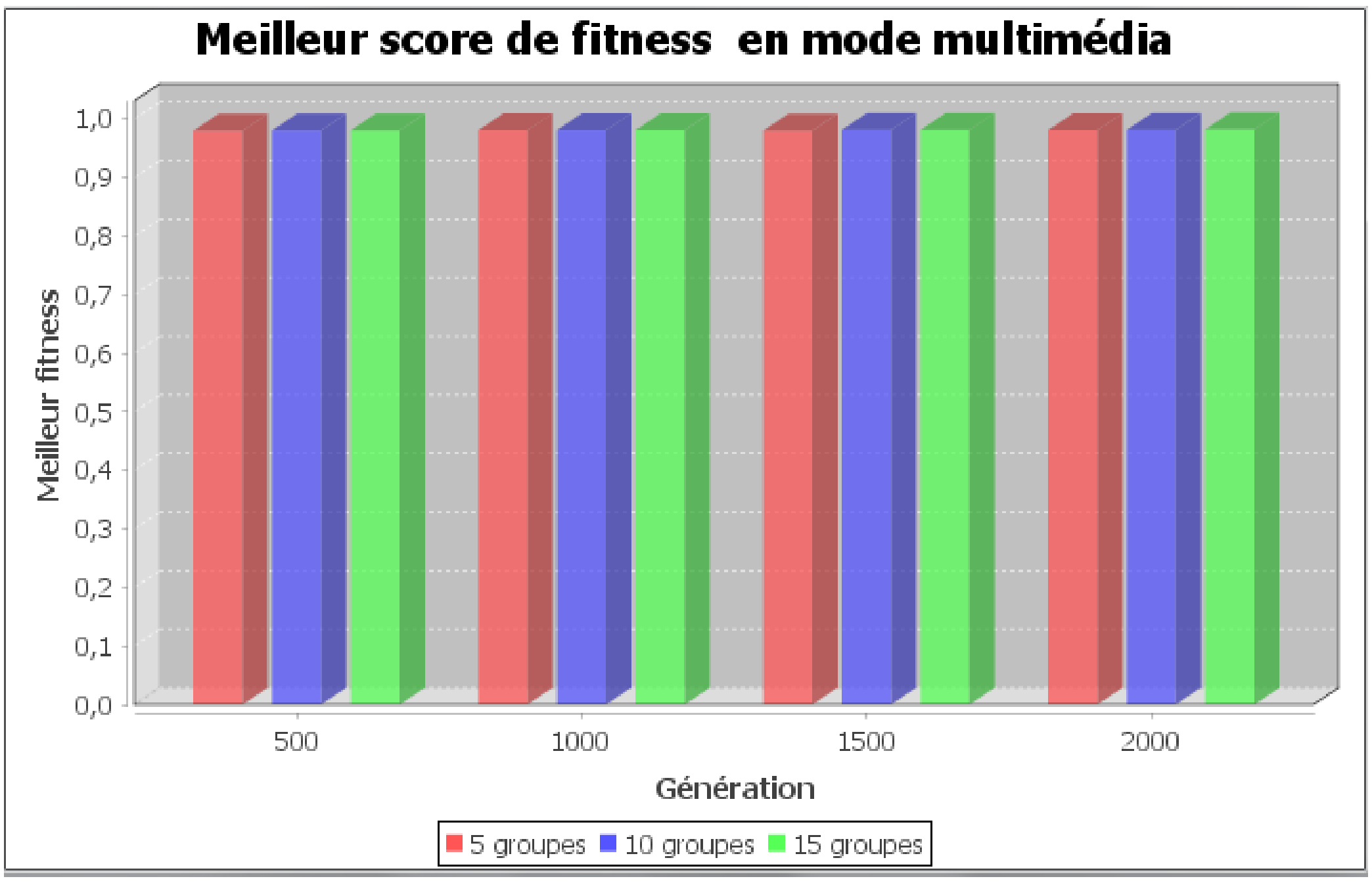

**Figure III.8:** Meilleur score de fitness en mode multimédia

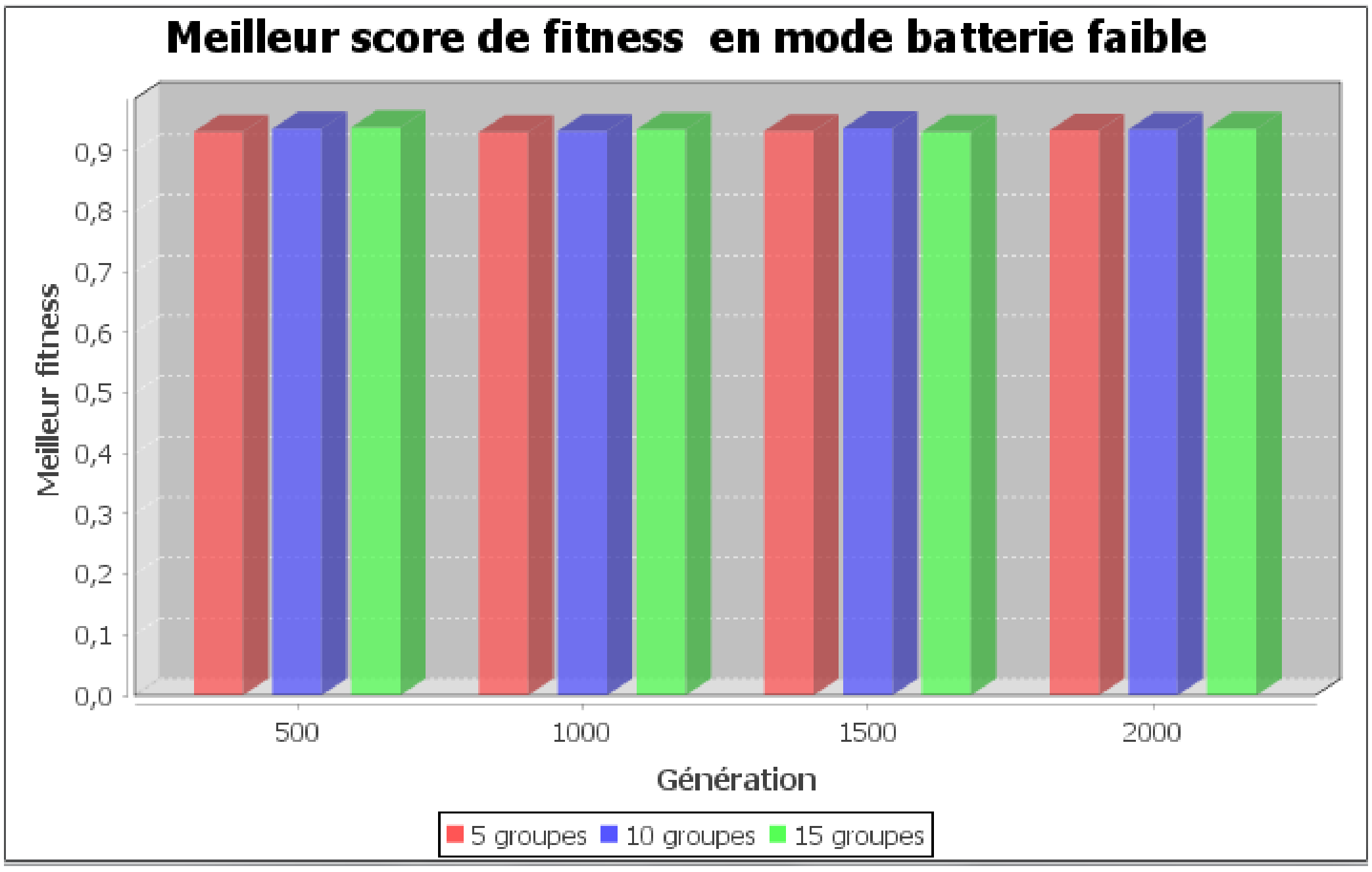

**Figure III.9 :** Meilleur score de fitness en mode batterie faible

## d. Temps d'exécution (ms)

Les différentes valeurs du tableau suivant représentent les résultats obtenus pour chaque mode (avec100 individus et 2000 générations).

| Mode / Nombre de canaux | Urgence | Multimédia | Batterie faible |
|---|---|---|---|
| 8 | 4506 | 3975 | 4377 |
| 16 | 9286 | 8081 | 7406 |
| 32 | 15806 | 12796 | 12396 |
| 64 | 28514 | 25285 | 25029 |
| 128 | 55335 | 42124 | 43625 |
| 256 | 137281 | 85912 | 111701 |
| 512 | 287150 | 187790 | 283331 |

**Tableau III.10:**Temps d'exécution en fonction du nombre de canaux dans les trois modes

La figure suivante représente le temps d'exécution en fonction du nombre de sous porteuses.

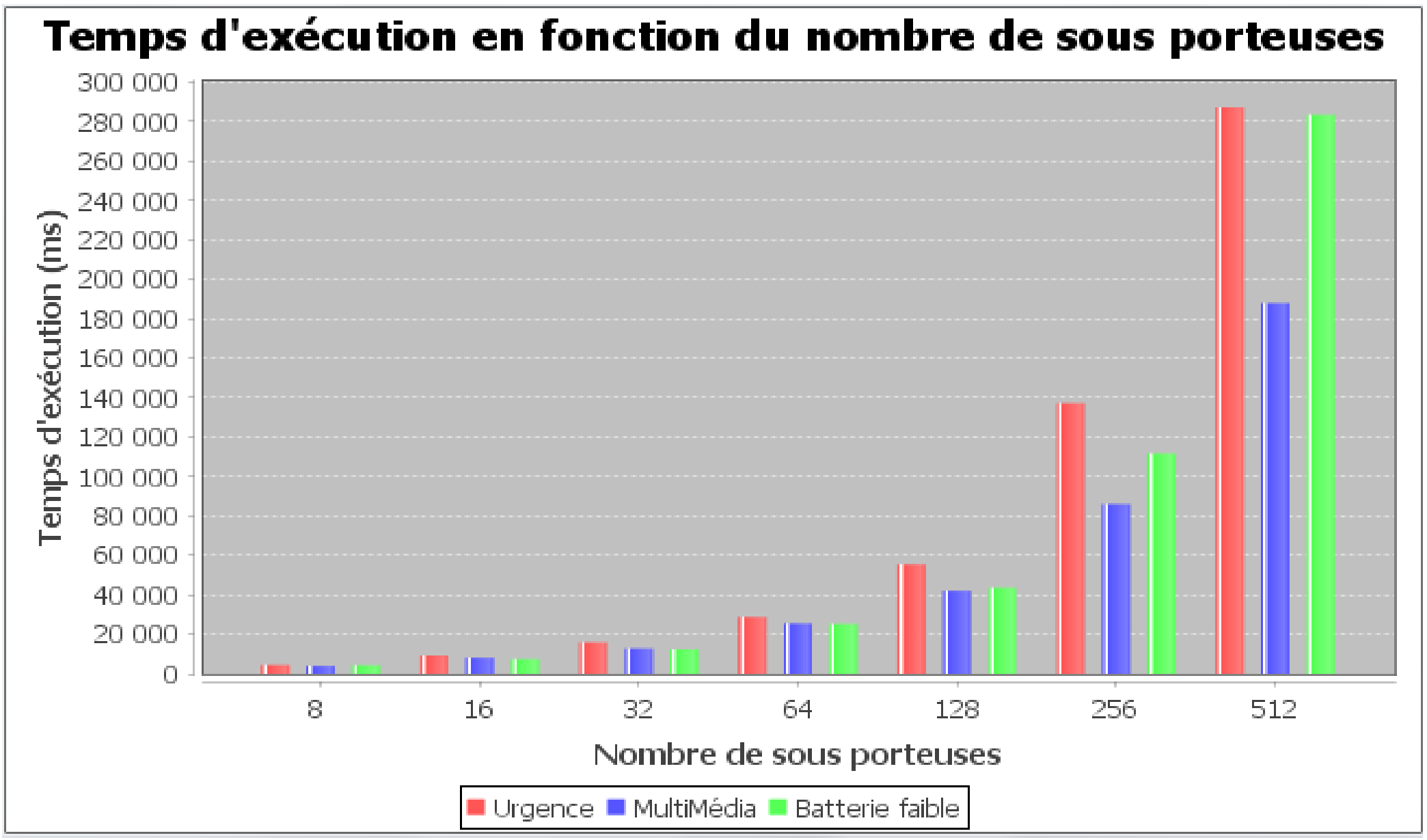

**Figure III.10 :** Temps d'exécution en fonction du nombre de sous porteuses

Les résultats précédents montrent que le temps d'exécution est devenu plus élevé en augmentant le nombre de canaux. Le temps le plus intéressant est réalisé avec 8 canaux dans le mode multimédia.

## III.10 Algorithme de saut de grenouille (SFLA) Vs Algorithme génétique (GA) en termes de fonction objectif

Dans ce rapport, nous avons utilisé la méthode d'optimisation stochastique SFLA pour optimiser la QoS dans un réseau de radio cognitive. Pour enrichir nos résultats, nous l'avons comparé avec les algorithmes génétiques qui ont été déjà appliqués dans [30].

Le tableau suivant représente les résultats obtenus en termes de meilleurs solution.

| Mode / Nombre de canaux | Urgence | | Multimédia | | Batterie faible | |
|---|---|---|---|---|---|---|
| | SFLA | GA | SFLA | GA | SFLA | GA |
| 8 | 0,88181 | 0.8198 | 0,98065 | 0.9701 | 0,93186 | 0.8955 |
| 16 | 0,84844 | 0.7983 | 0,98208 | 0.9749 | 0,93951 | 0.8936 |
| 32 | 0,82832 | 0.7926 | 0,98371 | 0.9701 | 0,94522 | 0.8996 |
| 64 | 0,81333 | 0.7983 | 0,98436 | 0.9695 | 0,95004 | 0.8945 |
| 128 | 0,79951 | 0.7853 | 0,98478 | 0.9690 | 0,95242 | 0.8962 |
| 256 | 0,79394 | 0.7823 | 0,98489 | 0.9668 | 0,95393 | 0.8948 |

**Tableau III.11:** Etude comparative entre SFLA et GA

La comparaison présentée dans le tableau précédent a mis l'accent sur la capacité de l'approche de saut de grenouille. Les résultats expérimentaux en termes de solution optimale globale ont montré que le SFLA est capable de garantir une meilleure solution dans les différents modes par rapport aux GA. En conséquence**,** le SFLA peut être un outil efficace pour résoudre les problèmes d'optimisation de QoS dans le contexte d'un réseau de radio cognitive.

## III.11 Conclusion

Dans ce chapitre, nous avons présenté les évaluations et tests réalisés dans le cadre de ce travail. Il s'agit de l'application de l'approche de grenouille sautant (SFLA) pour optimiser la QoS dans le cadre d'un réseau de radio cognitive. Les résultats obtenus de simulation ont montrés que cet algorithme fonctionne efficacement dans les différents modes d'application. Les résultats obtenus sont satisfaisants et l'algorithme présenté est mieux adapter que les algorithmes génétiques dans ce cadre d'application en termes de fonction objectif (fitness).

# Conclusion générale

Les réseaux de radio cognitive (RRC) sont des réseaux qui peuvent sentir leur environnement d'exploitation, d'adapter leur mise en œuvre et détecter les ressources disponibles comme les bandes de fréquence ou la présence d'utilisateur prioritaire pour atteindre les meilleures performances.

Nous avons focalisé nos recherche sur un des algorithmes métaheuristiques dit stochastiques destinées à résoudre des problèmes d'optimisation difficiles, cet algorithme est l'approche de saut de grenouille (Shuffled frog leaping algorithm).

Le SFLA est utilisé pour résoudre de nombreux problèmes. Il est relativement une bonne technique d'optimisation. Ainsi, il a l'avantage d'être facile à implémenter et ayant une vitesse rapide et une capacité d'optimisation globale.

C'est un algorithme robuste dans la détermination de la solution globale qui utilise les memplexes pour améliorer le taux de convergence et il se révèle d'être un algorithme très efficace.

En guise de perspectives, nous envisageons d'utiliser le couplage entre différents algorithmes métaheuristiques pour évaluer les performances de la radio cognitive à plus grande échelle selon l'exigence de l'utilisateur secondaire. Cette combinaison permet la majoration de la solution optimale dans un temps réduit.

# Références